\documentclass[aps,prd,onecolumn,showpacs,groupedaddress]{revtex4-2}
\usepackage{graphicx}
\usepackage{dcolumn}
\usepackage{bm}
\usepackage{color}
\usepackage{longtable}
\usepackage{supertabular}
\usepackage[T1]{fontenc}
\usepackage{epstopdf}
\usepackage{amsmath}
\usepackage{amssymb}
\usepackage{setspace}
\usepackage{multirow}
\usepackage{booktabs}
\usepackage[section]{placeins}
\usepackage{mathrsfs}
\usepackage{hyperref}

\makeatletter

\newcommand{\Rmnum}[1]{\expandafter\@slowromancap\romannumeral #1@}
\makeatother

\begin{document}
\title{A combinatorial algebraic approach for the modified second-generation time-delay interferometry}

\author{Zhang-Qi Wu\textsuperscript{1}}
\author{Pan-Pan Wang\textsuperscript{1}}\email[E-mail: ]{ppwang@hust.edu.cn}
\author{Wei-Liang Qian\textsuperscript{2,3,4}}\email[E-mail: ]{wlqian@usp.br}
\author{Cheng-Gang Shao\textsuperscript{1}}\email[E-mail: ]{cgshao@hust.edu.cn}

\affiliation{$^{1}$ MOE Key Laboratory of Fundamental Physical Quantities Measurement, Hubei Key Laboratory of Gravitation and Quantum Physics, PGMF, and School of Physics, Huazhong University of Science and Technology, Wuhan 430074, China}
\affiliation{$^{2}$ Escola de Engenharia de Lorena, Universidade de S\~ao Paulo, 12602-810, Lorena, SP, Brazil}
\affiliation{$^{3}$ Faculdade de Engenharia de Guaratinguet\'a, Universidade Estadual Paulista, 12516-410, Guaratinguet\'a, SP, Brazil}
\affiliation{$^{4}$ Center for Gravitation and Cosmology, College of Physical Science and Technology, Yangzhou University, Yangzhou 225009, China}

\begin{abstract}
We generalize the combinatorial algebraic approach first proposed by Dhurandhar {\it et al.} to construct various classes of modified second-generation time-delay interferometry (TDI) solutions.
The main idea behind the algorithm is to enumerate, in a given order, a specific type of commutator between two monomials defined by the products of particular time-displacement operators.
On the one hand, the above commutators can be systematically rewritten as the elements of a left ideal, defined by the l.h.s. of the relevant equation for the TDI solution.
On the other hand, these commutators are shown to vanish if we only keep up the first-order contributions regarding the rate of change of armlengths.
In other words, each commutator furnishes a valid TDI solution pertaining to the given type of modified second-generation combinations.
In this work, the original algorithm, which only involves time-delay operators, is extended by introducing the time-advance ones and then utilized to seek solutions of the Beacon, Relay, Monitor, Sagnac, and fully symmetric Sagnac types.
We discuss the relation between the present scheme's solutions and those obtained by the geometric TDI approach, a well-known method of exhaustion of virtual optical paths.
In particular, we report the results on novel Sagnac-inspired solutions that cannot be straightforwardly obtained using the geometric TDI algorithm. 
The average response functions, floor noise power spectral densities, and sensitivity functions are evaluated for the obtained solutions.

\end{abstract}

\date{Oct. 11th, 2022}

\maketitle

\section{Introduction}\label{sec1}

In terms of ground-based laser interferometry, the first ever direct detection of gravitational waves was accomplished by LIGO and Virgo collaborations in 2015~\cite{LIGO-1}.
The observation has been widely recognized as the inauguration of an era of gravitational-wave astronomy, as a new window to the universe had become available besides that via the electromagnetic spectrum.
Nonetheless, ground-based gravitational detections are constrained by a couple of crucial factors, such as the baseline length, Earth's seismic vibrations, and gravity-gradient noise.
To this end, the measurement is only feasible for the frequency band typically above 10Hz \cite{LIGO-2}. 
On the other hand, abundant wave sources, potentially for more robust and durable signals, are associated with a lower frequency band of 0.1mHz to 1Hz.
The latter is aimed at by the ongoing space-based gravitational wave detector projects, which include, notably, the LISA \cite{LISA-1}, TianQin \cite{TianQin-1}, Taiji \cite{Taiji-1}, and DECIGO \cite{DECIGO-1}. 
A space-borne detector typically consists of three identical spacecraft that form a giant, almost equilateral triangle configuration.
Encoded as Doppler frequency shifts, the information on the gravitational wave is embedded in the resultant beat notes between the laser beams exchanged among the spacecraft.
Unlike their ground-based counterpart, the relative frequency fluctuations primarily come from the laser's intrinsic phase instability.
In particular, the laser frequency noise cannot be simply canceled out using an equal-arm Michelson configuration.
This is because the distances between spacecraft vary in time according to the orbital dynamics, which gives rise to armlengths mismatch, where the rate of change of the armlength is between $5$ and $10\mathrm{m/s}$~\cite{LISA-1}.
Typically, the dominant laser frequency noise is about $7-8$ orders of magnitude higher than those from other sources~\cite{Overview-1}. 

In this regard, the TDI algorithm, first introduced by Tinto \emph{et al.} in 1999~\cite{TDI-1}, aims to construct a virtual equal-arm interferometer by a proper combination of the delayed science data streams in order to suppress the laser frequency noise~\cite{Overview-1}.
The existing TDI solutions can be divided into four categories: first-generation, modified first-generation, second-generation, and modified second-generation. 
The first-generation TDI~\cite{TDI1-1} approximates the entire triangular configuration as a rigid constellation without rotation, and subsequently, the detector arms are treated as static.
The modified first-generation TDI considers the Sagnac effect caused by a rigid rotation of the entire constellation~\cite{TDI1.5-1}.
For both cases, mathematically, the delay operations involved are commutative.
Therefore, the solution space is a polynomial ring $\mathscr{R}$ in three~\cite{TA-1} and six~\cite{TA-2} variables over the rational numbers.
Subsequently, the problem can be reformulated~\cite{TA-1, TA-2} to solve for the first {\it module of syzygies} of a relevant left ideal, whose generators can be obtained through the Groebner basis~\cite{groebner}.
The second-generation TDI further considers nonrigid rotation so that the armlengths vary independently but slowly in time~\cite{TDI2-1, Geo-sister}.
The residual noise is therefore evaluated as an expansion in terms of time-derivatives of the armlengths truncated at the second-order contributions.
The modified second-generation TDI discriminates between and enforces independent cancelation of distinct cyclic directions of the detector arms, leading to a more restrictive truncation scheme~\cite{Geo-sister}.
Owing to the non-commutative nature, the second and modified second-generation TDI solutions are not straightforward.
Vallisneri proposed the geometric TDI~\cite{Geo-TDI-1}, a method of exhaustion to seek TDI combinations by enumerating close trajectories in the space-time diagram.
The approach is intuitive because it can be shown that a geometric TDI solution corresponds to virtual equal-arm interferometry.
It has been extensively utilized in the literature to solve the second and modified second-generation TDI solutions~\cite{Olaf, Geo-sister}. 
Nonetheless, since the solution space of the geometric TDI method grows by $3^n$ where $n$ is the number of links, searching for TDI combinations is often computationally expensive.
Moreover, by definition, the algorithm demands that successive links must be ``connected''. 
Thus the solution space is somewhat restrictive.
In particular, the well-known fully symmetric Sagnac solutions of both generations lie beyond such a solution space.

Since the TDI algorithm was first proposed, it has flourished experimentally and theoretically in the literature. 
Regarding ground-based experiments, Vine \emph{et al.} verified the noise cancellation performance of the Sagnac TDI combination \cite{TE-1}, and more recently, Vinckier \emph{et al.} implemented the optical comb TDI scheme~\cite{TE-2} by using the acousto-optic modulators and optical combs. 
These results demonstrated TDI's viability in effectively suppressing laser frequency noise.
From the statistical inference perspective, Romano \emph{et al.}~\cite{R.2006} pointed out that the principal component analysis of the noise covariance matrix effectively furnishes a feasible TDI solution, and subsequently, further development of Bayesian TDI \cite{B.2021-1, B.2021-2} was proposed. 
Recently, Dhurandhar \emph{et al.} showed~\cite{TDI2022} a close relationship between two matrix-based TDI approaches that were independently proposed by Vallisneri~\cite{TM-1} \emph{et al.} and Tinto~\cite{TM-2} \emph{et al.}. 
In addition, efforts have been devoted to related topics such as eliminating residual clock noise using optical combs \cite{TO-1}. 
The clock noise cancellation scheme has also been generalized to the second-generation TDI combination by using sideband measurement \cite{TC-1, Clock-1}. 
Notably, Dhurandhar \emph{et al.} elaborated a combinatorial algebraic algorithm~\cite{D2010} for the modified second-generation TDI solutions in the case of one arm being dysfunctional.
In this simplified case, the polynomials that furnish a valid TDI solution are specific elements of a polynomial ring in four variables that corresponds to the kernel of the homomorphism
\begin{equation}
\varphi: \mathscr{R}^2\to \mathscr{R} ,
\end{equation}
namely, the first module of syzygies.
The proposed algorithm enumerates, in a given order, a specific type of commutator between two monomials defined by the products of particular time-displacement operators.
The authors showed that the commutators in question are the elements of the above left ideal $\mathscr{R}^2$.
Moreover, these commutators manifestly vanish if we only keep up the first-order contributions regarding the rate of change of armlengths.
In other words, these commutators correspond to valid TDI solutions.
It was shown how such solutions could be systematically constructed, and then the algorithm was applied to solve for the Michelson-type TDI combinations of the modified second-generation. 

Based on Dhurandhar {\it et al.}'s results~\cite{D2010}, the present paper further develops and explores the algorithm.
By inspecting the procedure, we argue that the interchange operation introduced in the original work is not an indispensable element.
Besides, the original algorithm, which only involves time-delay operators, is extended by introducing the time-advance ones.
As will be discussed below, the essence of the algorithm is the introduction of additional constraints to the TDI equation.
Consequently, the solution space simplifies so that the TDI solution can be constructed in terms of particular commutators.
Moreover, we show that the above generalizations expand the solution space, giving rise to a more flexible and robust algorithm.
Then, it is utilized to seek TDI solutions of the Beacon, Relay, Monitor, Sagnac, and fully symmetric Sagnac types.
In particular, we report a novel set of Sagnac-inspired solutions that cannot be straightforwardly derived using the geometric TDI. 
We also discuss the relationship between the present scheme's solutions and those obtained by the geometric TDI approach.
The average response functions, floor noise power spectral densities, and sensitivity functions are evaluated for the obtained solutions.

The remainder of the paper is organized as follows. 
In Sec.~\ref{sec2}, we briefly present the problem of the TDI algorithm, together with the notations and conventions used in this paper.
The combinatorial approach proposed by Dhurandhar {\it et al.} is discussed in Sec.~\ref{sec3}.
The original application to Michelson-type TDI solutions is revisited.
In Sec.~\ref{sec4}, the approach is elaborated further by including time-advance operators. 
Subsequently, the properties of the commutator are derived and discussed.
A more general version of the algorithm is formulated.
Subsequently, in Sec.~\ref{sec5}, the main results are presented by taking the Monitor type TDI solutions as an example.
We show that the nine sixteen-link modified second-generation TDI combinations, initially obtained by employing the geometric TDI, can be readily retrieved.
Furthermore, we apply the method to the case of fully symmetric Sagnac combinations.
Besides the well-known solutions in the literature, we present a few novel Sagnac-inspired solutions which cannot be straightforwardly obtained using the geometric TDI.
The average response functions, residual noise power spectral densities, and the sensitivity curves of the obtained novel solutions are evaluated.
Further discussions and concluding remarks are given in the last section. 
Some complementary derivations and discussions are relegated to the Appendices of the paper.
In Appendix~\ref{A}, we give proof of some of the mathematical relations utilized in the main text.
A few other classes of modified second-generation TDI solutions, namely, the Beacon, Relay, and Sagnac TDI combinations, are derived and presented in Appendix~\ref{B} using the method proposed in the present paper. 
In Appendix~\ref{C}, we enumerate the lower-order commutators that furnish the TDI solutions using the proposed algorithm.
We also show that two classes of higher-order TDI solutions can be {\it induced} using the lower-order ones, and the results are presented in Appendix~\ref{D}.
The explicit expressions of noise power spectral densities and the average response functions of the detector for the Sagnac-type combinations are given in Appendix~\ref{E}.

\section{Definitions, notations, and conventions of the TDI algorithm}\label{sec2}

\begin{figure}[h]
\includegraphics[width=0.50\textwidth]{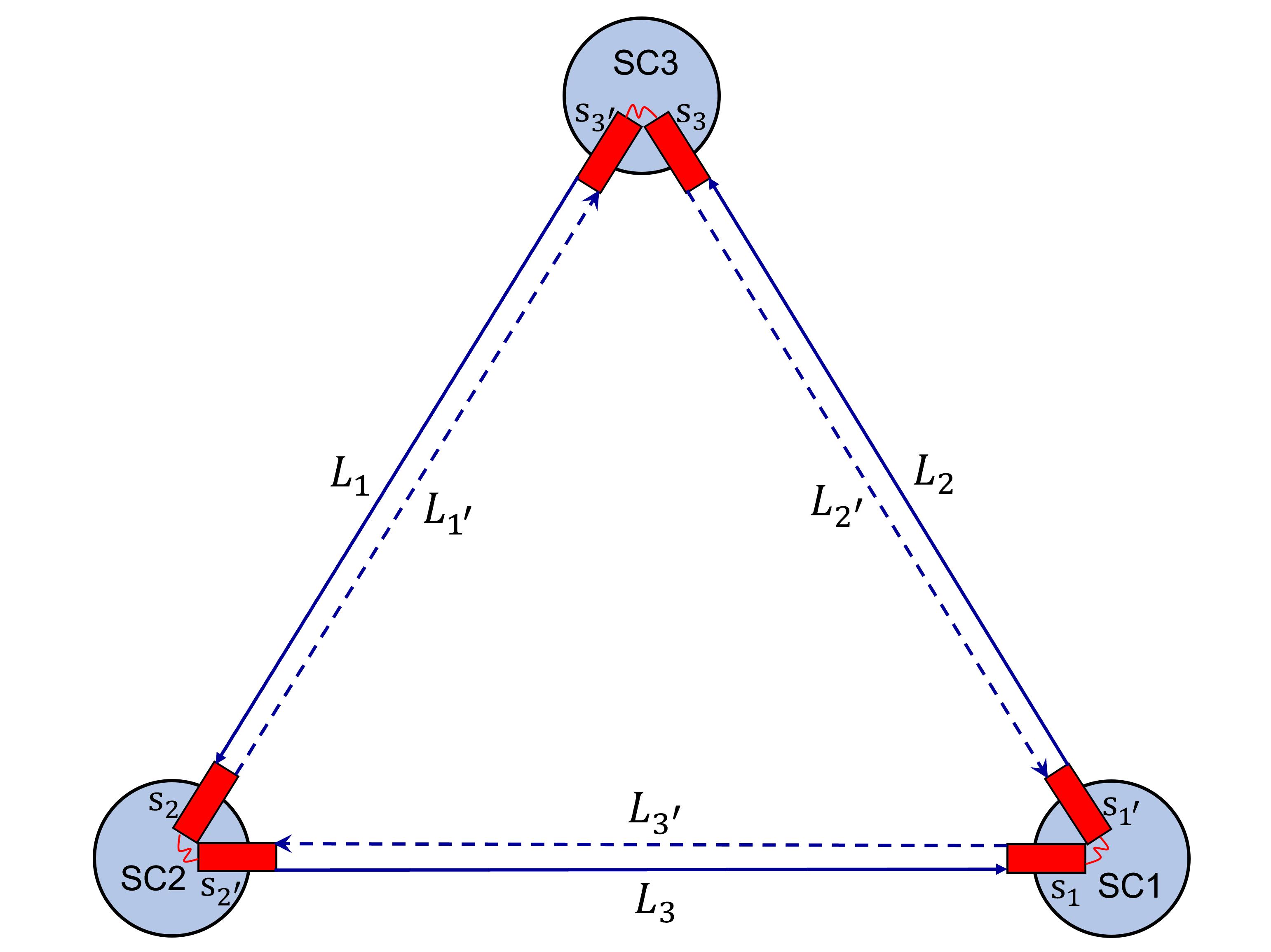}
\caption{\label{fig1} 
A schematic diagram of the three-spacecraft constellation for the space-borne gravitational wave detector.}
\end{figure}
\FloatBarrier

As illustrated in Fig.~\ref{fig1}, the experimental layout of a space-based gravitational wave detector consists of three identical spacecraft, denoted as SC$i$ (with $i=1, 2, 3$)~\cite{Overview-1}.
The armlengths sitting on the opposite side of SC$i$ are denoted as $L_i$ (and $L_{i'}$) in the counterclockwise (and clockwise) direction.
Two optical benches, labeled by $i$ and $i'$, are installed on each spacecraft.
The phasemeters of the optical benches perform the measurements of three types of data streams, namely, the science data streams $s_{i (i')}$, test mass data streams $\epsilon _{i (i')}$, and reference data streams $\tau _{i (i')}$.
The science data streams carry the essential information on the gravitational waves embedded in the beat notes between the laser beams from the distant and local spacecraft.
The test mass and reference data streams are formed by the interference between the two local laser beams from the adjacent optical benches $i$ and $i'$, whereas for the test mass data stream, one of the laser beams is reflected from the test mass.
By considering the laser frequency noise $p_{i (i')}$, optical bench motion noise $\dot{\vec{\Delta}}_{i (i')}$, test mass noise $\dot{\vec{\delta}}_{i, (i')}$, shot noise $N_{i (i')}^{opt}$ associated with the optical benches $i (i')$.
The relevant data streams recorded at the optical bench possess the following forms
	\begin{align} 
		s_{i} (t) &= D_{i-1} p_{{(i+1)}^{'}}(t) - p_{i} (t) + \nu_{(i+1)^{'}}[\vec{n}_{i-1}\cdot D_{i-1}\dot{\vec{ \Delta}}_{(i+1)^{'}}(t)+\vec{n}_{(i-1)^{'}}\cdot \dot{\vec{\Delta}}_{i}(t)] + H_{i} (t) + N_{i} ^{opt} (t),\notag\\
		 \epsilon _{i} (t)&= p_{{i}^{'}}(t) - p_{{i}}(t) - 2\nu_{i^{'}} [\vec{n}_{(i-1)^{'}} \cdot \dot{\vec{\delta}}_i (t)-\vec{n}_{(i-1)^{'}} \cdot \dot{\vec{\Delta}}_i (t)], \notag\\
		  \tau_i  (t)&= p_{i^{'}} (t) - p_i (t), 
	  \end{align} 
and
	\begin{align} 
		s_{i^{'}} (t) &= D_{{(i+1)}^{'}} p_{i-1}(t) - p_{i^{'}} (t) + \nu_{i-1}[\vec{n}_{i+1}\cdot \dot{\vec{\Delta}}_{i^{'}}(t)+\vec{n}_{(i+1)^{'}}\cdot D_{(i+1)^{'}}\dot{\vec{\Delta}}_{i-1}(t)]+ H_{i^{'}} (t) + N_{i^{'}} ^{opt} (t),\notag\\
		\epsilon _{i^{'}} (t)&= p_{{i}}(t) - p_{{i}^{'}}(t) - 2\nu_{i} [\vec{n}_{i+1} \cdot \dot{\vec{\delta}}_{i^{'}} (t)-\vec{n}_{i+1} \cdot \dot{\vec{\Delta}}_{i^{'}} (t)], \notag\\
		\tau_{{i}^{'}}  (t)&= p_i (t) - p_{i^{'}} (t),
	\end{align}
where $H_i (t), H_{i^{'}} (t)$ represent the gravitational wave signals, $D_{i (i')}$ are the time-delay operators along the related armlengths, $\nu_{i (i')}$ are the laser's frequency, and $\vec{n}_{i(i^{'})}$ are unit vectors along the armlengths.
Following the standard procedure of the TDI algorithm, the optical bench motion noise can be eliminated by introducing intermediate variables~\cite{TC-1}.
Subsequently, the two local lasers are effectively connected by intra-spacecraft phase locking, and the resultant observables read
	\begin{align}\label{eta}
		{\eta_i}(t) &= {H_i}(t) + {{D}_{i - 1}}{p_{i + 1}}(t) - {p_i}(t) + \nu_{(i+1)^{'}} {{\vec n}_{i - 1}}\left[ {{{D}_{i - 1}}{{\dot{\vec \delta }}_{(i + 1)'}}(t) - {{\dot {\vec \delta }}_i}(t)} \right] + N_i^{{opt}}(t) ,\notag\\
		{\eta_{i'}}(t) &= {H_{i'}}(t) + {{D}_{(i + 1)'}}{p_{i - 1}}(t) - {p_i}(t) + \nu_{i-1}{{\vec n}_{i + 1}} \cdot \left[ {{{\dot{\vec \delta }}_{i'}}(t) - {{D}_{(i + 1)'}}{{\dot{\vec{\delta} }}_{i - 1}}(t)} \right] + N_{i'}^{{opt}}(t) .
	\end{align}

In terms of the above observables, a valid TDI solution~\cite{Overview-1} is aimed to eliminate the three remaining independent laser frequency noise $p_i$ by a combination of the form
\begin{equation}\label{TDI}
	\mathrm{TDI}=\sum_{i=1,2,3} (q_{i} \eta_{i} + q_{i^{'}} \eta_{i'}) ,
\end{equation}
where $q_{i}$ and $q_{i'}$ are polynomials in the six time-delay operators.
By focusing on the laser frequency noise and explicitly demanding the coefficients before individual $p_{i}$ vanish, the above equation gives
\begin{equation}\label{TDIeq}
	\begin{aligned}
		&q_{1}+q_{1^{\prime}}-q_{2^{\prime}} {D}_{3^{\prime}}-q_{3} {D}_{2}=0,\\
		&q_{2}+q_{2^{\prime}}-q_{3^{\prime}} {D}_{1^{\prime}}-q_{1} {D}_{3}=0, \\
		&q_{3}+q_{3^{\prime}}-q_{1^{\prime}} {D}_{2^{\prime}}-q_{2} {D}_{1}=0.
	\end{aligned}
\end{equation}
Using two equations, one may choose to remove two of the six coefficients.
For instance, by eliminating $q_1$ and $q_2$, one finds
\begin{equation}\label{TDI4var}
	q_{3}(1-D_{231})+q_{1'}(D_{31}-D_{2'})+q_{2'}(D_{1}-D_{3'31})+q_{3'}(1-D_{1'1}) = 0.
\end{equation}

From the geometric TDI perspective~\cite{Geo-sister}, different generations of the TDI solutions can be classified by how the delayed laser frequency noise, as an expansion using time derivatives of the armlengths, are truncated.
To be specific, for the first-generation TDI, one considers
\begin{equation}\label{LTDI1}
	L_{i}(t)=L_{i}=\mathrm{const.},\ \ L_{i}=L_{i^{'}} .
\end{equation}
For the modified first-generation TDI, one distinguishes armlengths of different cyclic orders, namely,
\begin{equation}\label{LTDI1m}
	L_{i}(t)=L_{i}=\mathrm{const.},\ \ L_{i}\neq L_{i^{'}} .
\end{equation}

Following the literature~\cite{TDI2-1}, one refers to the case where
\begin{equation}\label{LTDI2}
	L_{i}(t)=L_{i}+t\dot{L}_{i}, \ \ L_{i}\neq L_{i^{'}}, \ \ \dot{L}_{i}=\dot{L}_{i^{'}}
\end{equation}
as the second-generation TDI.
Lastly, the solutions satisfying
\begin{equation}\label{LTDI2m}
	L_{i}(t)=L_{i}+t\dot{L}_{i}, \ \ L_{i}\neq L_{i^{'}}, \ \ \dot{L}_{i}\neq \dot{L}_{i^{'}} ,
\end{equation}
as investigated by the studies~\cite{Geo-sister} will be dubbed as the {\it modified} second-generation TDI~\footnote{It is noted that the last case was referred to as the second-generation TDI in some literature~\cite{2020CQG}.}.

For the first-generation TDI defined in Eqs.~\eqref{LTDI1} and~\eqref{LTDI1m}, the polynomials $q_{i (i')}$ form a commutative ring which allows us to solve for the TDI combination using the Groebner basis~\cite{groebner}. 
On the other hand, the delay operators in Eq.~\eqref{TDIeq} can no longer be viewed as commutative in the context of the second generation TDI defined in Eqs.~\eqref{LTDI2} and~\eqref{LTDI2m}.
To be more precise, the solution space of Eq.~\eqref{TDIeq} is a left module over a non-commutative ring $\mathscr{R} = \mathbb{Q} (D_{1},D_{2},D_{3},D_{1^{'}},D_{2^{'}},D_{3^{'}})$, where $\mathbb{Q}$ is the field of rational numbers.
However, the non-commutative nature of the problem implies that it is not straightforward to solve for the general and exhaustive form of the TDI combinations using an algebraic approach.
In the next section, the properties of the commutators between polynomials of the delay operator will be studied, which eventually leads to a combinatorial algebraic approach.

\section{The original combinatorial algebraic algorithm}\label{sec3}

\subsection{A specific type of commutators and the vanishing condition}\label{sec3.1}

By applying the delay operators $D_{i}$ and $D_{i^{'}}$ on an arbitrary time-dependent variable $\phi (t)$, we have
\begin{equation}\label{D12def}
	\begin{aligned}
		D_{i} \phi(t) &=\phi(t-L_{i} (t)),\\
		D_{ji}\phi (t) & = \phi (t-L_{j}(t)-L_{i} (t-L_{j} (t))) ,
	\end{aligned}
\end{equation}
where, for convenience, one has denoted successive applications of the time-delay operators by
\begin{equation}
	D_{j} D_{i}\phi (t) \equiv D_{ji} \phi (t).
\end{equation}

By making use of the expansion
\begin{equation}
	L_i (t)= L_i+t\dot{L}_i+\frac{1}{2}t^2\ddot{L}_{i}+\cdots ,
\end{equation}
where the speed of light in the vacuum is taken as unit $c=1$, to the first order, Eqs.~\eqref{D12def} give,
\begin{equation}\label{1D}
	D_{i} \phi (t) \simeq \phi (t-L_{i}) - \dot{\phi}(t-L_{i}) t \dot{L}_{i},
\end{equation}
and
\begin{equation}\label{1D22}
	\begin{aligned}
		D_{ji} \phi(t) &\simeq \phi(t-L_{j} (t) - L_{i} (t) + \dot{L}_{i} L_{j}(t))\\ 
		&\simeq \phi(t-L_{i}-L_{j}) + \dot{\phi}(t-L_{i}-L_{j})\dot{L}_{i} L_{j}\\
		&-\dot{\phi}(t-L_{i}-L_{j})t(\dot{L}_{i}+\dot{L}_{j}) .
	\end{aligned}
\end{equation}

It is straightforward to generalize the above results to higher orders.
For instance, by applying three time-delay operators, one finds
\begin{equation}\label{1D33}
	\begin{aligned}
	D_{kji} \phi (t) &\simeq \phi (t-L_{i}-L_{j}-L_{k}) \\
	&+ \dot{\phi}(t-L_{i}-L_{j}-L_{k}) \big(\dot{L}_{i} (L_{j} + L_{k})+\dot{L}_{j} L_{k}\big)\\
	&-\dot{\phi}(t-L_{i}-L_{j}-L_{k})t(\dot{L}_{i}+\dot{L}_{j}+\dot{L}_{k}).
	\end{aligned}
\end{equation}

It was first pointed out in~\cite{D2010} that many TDI combinations can be attributed to a specific type of commutator whose form is related to the residual of the laser frequency noise.
For convenience, in what follows, we will use the subscript $x$ to denote the first element of a commutator, and the subscript $y$ will be reserved for the second element of the commutator.
For instance, a commutator will be formally written as $[D_{x_{1} x_{2} \dots x_{n}},D_{y_{1} y_{2} \dots y_{n}}]$.
Let us denote the commutator associated with the residual laser noise of the TDI solution as
\begin{equation}
	\mathrm{TDI}^{p}=[D_{x_{1} x_{2} \dots x_{n}},D_{y_{1} y_{2} \dots y_{n}}] p_{i} (t),
\end{equation}
where $x_{i}$ and $y_{i}$ are one of time-delay operators $D_{i}$ or $D_{i^{'}}$.
 
We illustrate the above statement with two examples.
Let us take the modified first-generation Michelson combination $X_{1}$ as the first example.
We have
\begin{equation}
	\begin{aligned}\label{MX1}
	{X}_{1}(t)&=\eta_{1}+D_{3} \eta_{2^{\prime}}+D_{33^{\prime}} \eta_{1^{\prime}}+D_{33^{\prime} 2^{\prime}} \eta_{3}\\
	&- \left(\eta_{1^{\prime}}+D_{2^{\prime}} \eta_{3}+D_{2^{\prime} 2} \eta_{1}+D_{2^{\prime} 23} \eta_{2^{\prime}}\right) .
	\end{aligned}
\end{equation}
It is straightforward to show that the residual of the laser frequency noise is
\begin{equation}\label{r1}
	 X^{p}_1(t)=[D_{3 3^{'}},D_{2^{'} 2}] p_{1} (t).
\end{equation}
For the second example, we consider the modified second-generation Michelson combination $X_{2}$, 
\begin{equation}\label{M2}
	\begin{aligned}
		X_{2}(t)&=\eta_{1}+{D}_{3} \eta_{2^{\prime}}+{D}_{33^{\prime}} \eta_{1^{\prime}}+{D}_{33^{\prime} 2^{\prime}} \eta_{3}+{D}_{33^{\prime} 2^{\prime} 2} \eta_{1^{\prime}} \\
		&+{D}_{33^{\prime} 2^{\prime} 22^{\prime}} \eta_{3}+{D}_{33^{\prime} 2^{\prime} 22^{\prime} 2} \eta_{1}+{D}_{33^{\prime} 2^{\prime} 22^{\prime} 23} \eta_{2^{\prime}} \\
		&-\left(\eta_{1^{\prime}}+{D}_{2^{\prime}} \eta_{3}+{D}_{2^{\prime} 2} \eta_{1}+{D}_{2^{\prime} 23} \eta_{2^{\prime}}+{D}_{2^{\prime} 233^{\prime}} \eta_{1}\right. \\
		&\left.+{D}_{2^{\prime} 233^{\prime} 3} \eta_{2^{\prime}}+{D}_{2^{\prime} 233^{\prime} 33^{\prime}} \eta_{1^{\prime}}+{D}_{2^{\prime}233^{\prime} 332^{\prime}} \eta_{3}\right),
	\end{aligned}
\end{equation}
the residual reads
\begin{equation}\label{r2}
	 X^{p}_2(t)=[D_{33^{'}2^{'}2},D_{2^{'}233^{'}}]p_1 (t).
\end{equation}

From the two Michelson combinations given by Eqs.~\eqref{r1} and~\eqref{r2}, it is observed that the residuals can always be written as commutators composed of the products of two components $D_{33^{'}}$ and $D_{2^{'}2}$. 
This indicates that one may construct TDI solutions from specific commutators satisfying some rules.
Indeed, modified second-generation Michelson-type combinations of higher orders can be obtained from the following residuals:
\begin{equation}
[D_{33^{'}2^{'}22^{'}233^{'}},D_{2^{'}233^{'}33^{'}2^{'}2}], \nonumber
\end{equation}
\begin{equation}
[D_{33^{'}33^{'}2^{'}22^{'}2},D_{2^{'}22^{'}233^{'}33^{'}}], \nonumber
\end{equation}
and
\begin{equation}
[D_{33^{'}2^{'}233^{'}2^{'}2},D_{2^{'}233^{'}2^{'}233^{'}}]. \nonumber
\end{equation}

A systematic approach based on the above speculations was first proposed in~\cite{D2010}.
It consists of two crucial building blocks.
First, one identifies a specific type of commutator corresponding to vanishing residual laser noise.
Second, these commutators can be mapped to the TDI solution characterized by the corresponding coefficients $q_{i (i')}$.
The first point will be addressed shortly, and Sec.~\ref{sec3.2} will be devoted to the second point.

In~\cite{D2009}, the following relation was established
\begin{equation}\label{sys1}
	\begin{gathered}
		{\left[D_{x_{1} x_{2} \ldots x_{n}}, D_{y_{1} y_{2} \ldots y_{n}}\right] \phi(t)} 
		=\left(\sum_{k=1}^{n} L_{x_{k}} \sum_{m=1}^{n} \dot{L}_{y_{m}}-\sum_{m=1}^{n} L_{y_{m}} \sum_{k=1}^{n} \dot{L}_{x_{k}}\right) 
		\dot{\phi}\left(t-\sum_{k=1}^{n} L_{x_{k}}-\sum_{m=1}^{n} L_{y_{m}}\right).
	\end{gathered}
\end{equation}
For the lower order cases, it can be readily verified using Eqs.~\eqref{1D}-\eqref{1D33}.
A general proof of Eq.~\eqref{sys1} is relegated to Appendix~\ref{A}. 

Apparently, the first factor on the r.h.s. of Eq.~\eqref{sys1} vanishes as long as
\begin{equation}\label{permuXY}
	y_i=x_{\pi(i)} .
\end{equation}
where $\pi\in \mathcal{S}_n$ is an arbitrary element of the permutation group of degree $n$. 
In other words, Eq.~\eqref{sys1} vanishes for a specific type of commutator characterized by a given number of indices and specific permutations.
Moreover, if one can show that, under certain circumstances, such commutator can be recognized as, for instance, the l.h.s. of Eq.~\eqref{TDI4var}, it furnishes a valid TDI solution.
The latter proposition will be explored in the following subsection.
For the remainder of the present subsection, however, we elaborate further on the converse proposition by revisiting the Michelson combinations Eq.~\eqref{M2} discussed above.

As a matter of fact, this class of solutions has been extensively explored in~\cite{D2010}, referred to as one-arm dysfunctional ones.
Without loss of generality, one assumes that the communications through the armlength connecting SC2 and SC3 are interrupted.
In other words, we have $\eta_{2}=\eta_{{3}^{'}}=0$, or equivalently,
\begin{equation}\label{MC}
	q_{2}=q_{3^{'}}=0 .
\end{equation}
Substituting Eq.~\eqref{MC} into Eq.~\eqref{TDIeq}, we have
\begin{subequations}\label{abc}
	\begin{align}
		&q_{1}+q_{1^{\prime}}-q_{2^{\prime}} {D}_{3^{\prime}}-q_{3} {D}_{2}=0, \label{a}\\
		&q_{2^{\prime}}-q_{1} {D}_{3}=0, \label{b}\\
		&q_{3}-q_{1^{\prime}} {D}_{2^{\prime}}=0\label{c} .
	\end{align}
\end{subequations}
Eliminating Eqs.~\eqref{b} and \eqref{c} by substituting the forms of $q_{2'}$ and $q_3$ into Eq.~\eqref{a}, one finds the desired equation
\begin{equation}\label{MTDI}
	q_{1}(1-D_{33^{'}})+q_{1^{'}}(1-D_{2^{'}2})=0.
\end{equation}

We note that the above elimination process algebraically ensures that the residual laser noise is entirely governed by $p_{1}$. 
If one had chosen to eliminate two other equations, the residual would have been determined by other laser noise.
The difference, however, is irrelevant from a modified second-generation TDI perspective. 
Also, by adding distinct solutions, one may construct an optimal solution that minimizes a specific target quantity.

By observing the specific form of Eq.~\eqref{MTDI}, it is also apparent that $D_{33^{'}}$ and $D_{2^{'}2}$ are the elementary units to furnish any solution.
For convenience, we denote $a=D_{33^{'}}$ and $b=D_{2^{'}2}$, and Eq.~\eqref{MTDI} simplifies to read
\begin{equation}\label{nMTDI}
	q_{1}(1-a)+q_{1^{'}}(1-b)=0.
\end{equation}
Also, the solution Eq.~\eqref{M2} corresponds to the coefficients $q_{1}$ and $q_{1^{'}}$
\begin{equation}\label{M16}
	\begin{aligned}
	q_{1} &= 1-b-ba+ab^2,\\
	q_{1^{'}} &= - (1-a-ab+ba^2) ,
	\end{aligned}
\end{equation}
and the residual Eq.~\eqref{r2} reads 
\begin{equation}\label{M16V2}
[ba, ab]p_1(t).
\end{equation} 

Following~\cite{D2010}, one can interpret the above residual as that of a cancelation taking place between the two terms on the l.h.s. of Eq.~\eqref{nMTDI} in an order-by-order fashion.
In other words, this motivated a procedure regarding how each term of the two expressions given by Eqs.~\eqref{M16} is written down.
As all the lower order terms are canceled out identically, the procedure continues until the last two remaining terms of the highest order {\it coincidently} give rise to a commutator, which possesses the form of the l.h.s. of Eq.~\eqref{sys1} while satisfying Eq.~\eqref{permuXY}.
Subsequently, the commutator vanishes, reassuring that Eqs.~\eqref{M2} or~\eqref{M16} is indeed a valid TDI solution.

Dhurandhar {\it et al.} introduced an interchange operator $\mathcal{I}$ between the symbols $a$ and $b$~\cite{D2010}.
To be specific, for a given monomial $s$ composed of $a$ and $b$, the string $t = \mathcal{I}(s)$ is obtained by replacing $a$ with $b$, and meanwhile substituting original $b$ by $a$, in $s$.
For instance, $\mathcal{I}(abba) = baab$.
The authors pointed out that for the one-arm dysfunctional case, namely, the Michelson combinations, the solutions possess the form $[s, \mathcal{I}(s)]$.
Moreover, in accordance with the l.h.s. of Eq.~\eqref{sys1}, if one has $n$ instances of $a$ in the string $s$, the count of $b$'s must be the same.
Therefore, the length of the string $s$ is $2n$, and the commutator consists of strings of length $4n$.

\subsection{TDI solutions derived from the commutators}\label{sec3.2}

The discussions elaborated by the end of the last subsection indicate that one may derive TDI solutions by exhaustively enumerating all possible commutators.  
To be specific, for $n=1$, the only commutator is $[ab, ba]$, which corresponds to Eq.~\eqref{M16V2}.
For $n=2$, one has three options: $[a^{2}b^{2},b^{2}a^{2}]$, $[abab,baba]$, and $[ab^2a,ba^2b]$. 
For an arbitrary integer number $n$, there are $\frac{(2n)!}{2n!n!}$ relevant combinations of the form
\begin{equation}\label{residual}
\Delta=[s_{2n},\mathcal{I}(s_{2n})]=s_{2n}\mathcal{I}(s_{2n})-\mathcal{I}(s_{2n})s_{2n},
\end{equation}
where $s_{2n}$ is an monomial comprised of $n$ instances of $a$ and $b$. 
Also, $q_{1'} = -\mathcal{I}(q_{1})$,  which implies that we only need the explicit form of $q_{1}$ for the solution of Eq.~\eqref{nMTDI}. 
The remaining coefficients $q_{3}, q_{2^{'}}$ are subsequently obtained by Eqs.~\eqref{b} and~\eqref{c}.

\begin{figure*}
\centering
\includegraphics[width=1\textwidth]{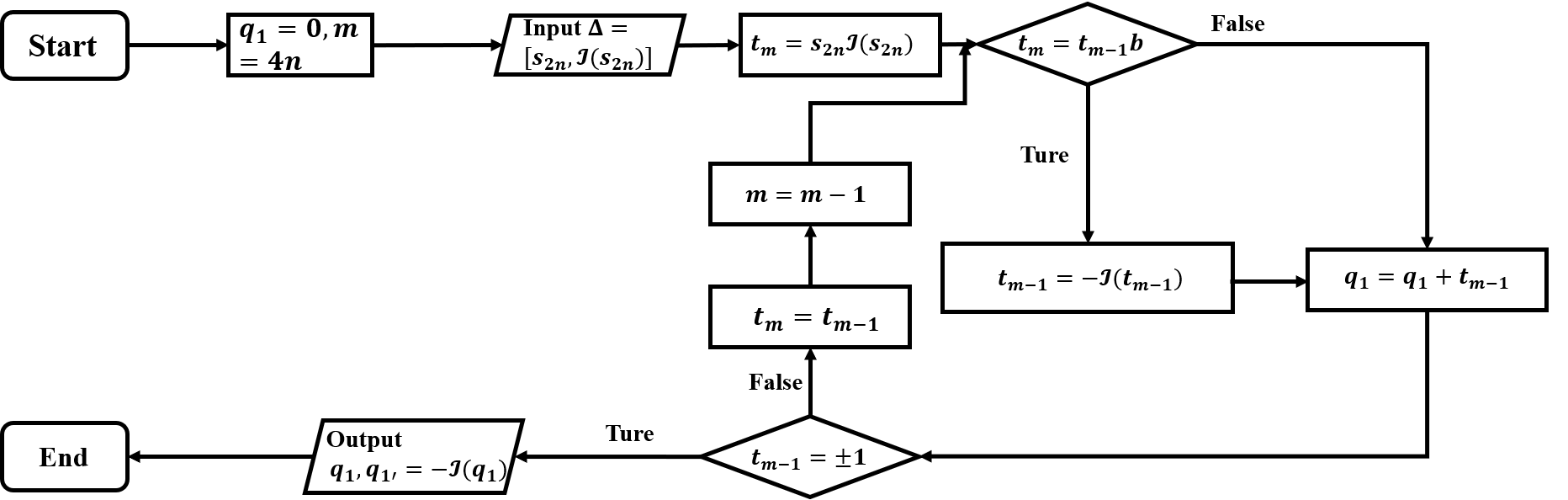}
\caption{\label{process1} 
The control-flow diagram of the algorithm proposed in~\cite{D2010} for Michelson TDI combinations.}
\end{figure*}

Based on the above arguments, an algorithm was proposed~\cite{D2010} to derive TDI solutions for the one-arm dysfunctional case systematically.
The idea of the algorithm is first to enumerate all possible commutators in question, then deduce the polynomial coefficients $q_{i (i')}$ of the TDI combinations from a given commutator. 
The latter process is reiterated below, illustrated by the control-flow diagram Fig.~\ref{process1}.
\begin{enumerate}
\item Consider the first term in Eq.~\eqref{residual}, $t_{4n}\equiv s_{2n} \mathcal{I}(s_{2n})$, it is a monomial of degree $4n$.
It is noted that $t_{4n}$ ends in either $a$ or $b$, that is, $t_{4n}=t_{4n-1}a$ or $t_{4n}=t_{4n-1}b$. 
\begin{itemize}
\item If $t_{4n}=t_{4n-1}a$, let $t_{4n-1}=t_{4n-1}$; 
\item If $t_{4n}=t_{4n-1}b$, let $t_{4n-1}=-\mathcal{I}(t_{4n-1})$.
\end{itemize}
\item Repeat the above procedure for $t_{4n-1}$ until the degree of the monomial vanishes, namely, $t_{0}$.
We note that $t_{0}=\pm 1$.
\item The TDI coefficients are obtained by summing up $4n$ monomials: $q_{1}=\sum_{k=0}^{4n-1} t_{k}$ and $q_{1^{'}}=-\mathcal{I}(q_{1})$.
\end{enumerate}
As an example, for the case of $[ab,ba]$ which $n=1$.
The first term is $t_{4}=ab^{2}a$, then we have $t_{3}=ab^2$, $t_{2}=-\mathcal{I}(ab)=-ba$, $t_{1}=-b$, and $t_{0}=1$.
It is readily verified that Eq.~\eqref{M16} is retrieved by summing up the above $4$ monomials.

\section{The extended combinatorial algebraic algorithm}\label{sec4}

In the last section, we revisited an existing algorithm dedicated to deriving the modified second-generation Michelson TDI combinations using a specific type of commutator. 
The present section extends the above approach to other classes of TDI solutions.
The generalization is essentially based on the following considerations:
\begin{itemize}
\item The algorithm proposed in~\cite{D2010} does not depend on the specific forms of $a$ and $b$, and it is viable as long as one encounters an equation of the form Eq.~\eqref{nMTDI}.
\item Algebraically, to derive something similar to Eq.~\eqref{nMTDI}, one only needs to introduce two additional constraints (e.g. Eq.~\eqref{MC}) into Eq.~\eqref{TDIeq}.
\item The introduction of the interchange operation $\mathcal{I}$ is not an obligation. One only needs the condition Eq.~\eqref{permuXY} for commutator Eq.~\eqref{sys1} to vanish.
\item The number of variables of the non-commutative ring $\mathscr{R}$ can be expanded to include time-advance operators, which leads to a generalized commutation relation.
\end{itemize}
These points give rise to an extended version of the algorithm, which can be used to derive TDI solutions of other classes, such as Beacon, Relay, Monitor, Sagnac, and full symmetric Sagnac ones.
Moreover, one encounters novel Sagnac-inspired solutions that cannot be straightforwardly derived using the geometric TDI.

In the remainder of the present section, we will elaborate on the general algorithm.
In particular, the generalized communtation relation will be discussed in Sec.~\ref{sec4.1}.
Explicit examples will be explored in Sec.~\ref{sec5}.

\subsection{The generalized commutator}\label{sec4.1}

By observing other classes of TDI combinations well-known in the literature, it is noted that the polynomial coefficients often contain time-advance operators $D_{\bar{i} (\bar{i'})}$.
They are the inverse of the corresponding time-delay operators, satisfying
\begin{equation}\label{DbarDef}
	D_{\bar{i}} D_{i} =  D_{i} D_{\bar{i}} = \mathscr{I} ,
\end{equation}
where $\mathscr{I}$ is the identity operator.
When a time-advance operator is applied to a time-dependent variable $\phi(t)$, Eq.~\eqref{DbarDef} implies, to the first order,
\begin{equation}
D_{\bar{i}} \phi(t) = \phi (t+L_{i}(t+L_{i})) ,
\end{equation}
which gives
\begin{equation}\label{1Dinver}
	\begin{aligned}
		{D}_{\overline{i}} \phi(t) & \simeq \phi\left({t}+{L}_{{i}}\right)+\dot{\phi}\left(t+L_{i}\right)\left(t+L_{i}\right) \dot{L}_{i} \\
		&=\phi\left({t}+{L}_{{i}}\right)+\dot{\phi}\left(t+L_{i}\right) t \dot{L}_{i}+\dot{\phi}\left(t+L_{i}\right) L_{i} \dot{L}_{i}.
	\end{aligned}
\end{equation}

By comparing Eq.~\eqref{1Dinver} against Eq.~\eqref{1D}, one observes that in the argument of the first and second terms, there is an extra $(-1)$ factor for $L_{i}$ and $\dot{L}_{i}$.
Besides, the last term on the second line of Eq.~\eqref{1Dinver} gives rise to a novel contribution. 

Now, we proceed to show that Eq.~\eqref{sys1} continues to be valid when the time-advance operator is included.
This can be done by appropriately modifying the proof given in Appendix~\ref{A}, as follows.
We assume that some operators in the l.h.s. of Eq.~\eqref{sys1} are time-advance ones and introduce modifications to the existing derivations.
Without loss of generality, let us assume there are $l$ instances of time-advance operators in $D_{x_{1} x_{2} \dots x_{n}}$ and $m$ instances in $D_{y_{1} y_{2} \dots y_{n}}$.
To be specific, we denote the subscripts corresponding to these operators by $\lambda_i$ with ($i=1, \cdots, l$) and $\gamma_j$ with ($j=1, \cdots, m$).
Also, the additional $(-1)$ factors in Eq.~\eqref{1Dinver} can be accounted for by the factor
\begin{equation}
\begin{aligned}
\delta x_k &= \left\{  \begin{matrix}-1&&\mathrm{if}\ k=\lambda_i\ \mathrm{for\ any}\ i=1,\cdots,l\\+1&&\mathrm{otherwise}\end{matrix} \right. ,\\
\delta y_{k'} &= \left\{  \begin{matrix}-1&&\mathrm{if}\ k'=\gamma_j\ \mathrm{for\ any}\ j=1,\cdots,m\\+1&&\mathrm{otherwise}\end{matrix} \right. .
\end{aligned}
\end{equation}
By substituting Eq.~\eqref{1Dinver} into $\left[D_{x_{1} x_{2} \ldots x_{n}}, D_{y_{1} y_{2} \ldots y_{n}}\right]$, the novel contributions owing to the third term in Eq.~\eqref{1Dinver} read
\begin{equation}
\begin{aligned}
\left(\sum_{i=1}^{l} \dot{L}_{x_{\lambda_i}} L_{x_{\lambda_i}}+\sum_{j=1}^{m} \dot{L}_{y_{\gamma_j}} L_{y_{\gamma_j}}-\sum_{i=1}^{l} \dot{L}_{x_{\lambda_i}} L_{x_{\lambda_i}}-\sum_{j=1}^{m} \dot{L}_{y_{\gamma_j}} L_{y_{\gamma_j}}\right)\dot{\phi}\left(t-\sum_{k=1}^{n} \delta_{x_{k}}L_{x_{k}}-\sum_{k'=1}^{n} \delta_{y_{k'}}L_{y_{k'}}\right)=0 ,
\end{aligned}
\end{equation}
which manifestly vanishes.
Therefore, by putting all the pieces together, we arrive at the following generalized form of Eq.~\eqref{sys1}
\begin{equation}\label{sys2}
	\begin{gathered}
		{\left[D_{x_{1} x_{2} \ldots x_{n}}, D_{y_{1} y_{2} \ldots y_{n}}\right] \phi(t)} 
		=\left(\sum_{k=1}^{n} \delta_{x_{k}} L_{x_{k}} \sum_{k^{'}=1}^{n} \delta_{y_{k^{'}}}\dot{L}_{y_{k^{'}}}-\sum_{k^{'}=1}^{n} \delta_{y_{k^{'}}} L_{y_{k^{'}}} \sum_{k=1}^{n} \delta_{x_{k}} \dot{L}_{x_{k}}\right) 
		\dot{\phi}\left(t-\sum_{k=1}^{n} \delta_{x_{k}}L_{x_{k}}-\sum_{k'=1}^{n} \delta_{y_{k'}}L_{y_{k'}}\right) .
	\end{gathered}
\end{equation}

It is now apparent that the vanishing condition Eq.~\eqref{permuXY} remains unchanged when the time-advance operators are considered.
By using a broadened choice of variables provided by the generalized commutation relation Eq.~\eqref{sys2}, in the following subsection, we elaborate on an extended combinatorial algebraic approach for the modified second-generation TDI.

\subsection{The extended combinatorial algorithm}\label{sec4.2}

We start by discussing an explicit example not embraced by the original algorithm.
To be specific, we consider alternative Michelson combination~\cite{Geo-TDI-1, Geo-sister} given by
\begin{equation}
	\begin{aligned}
		{X}_{2}(t) &=\left[\eta_{1}+D_{3} \eta_{2^{\prime}}+D_{33^{\prime}} \eta_{1^{\prime}}+D_{33^{\prime} 2^{\prime}} \eta_{3}+D_{33^{\prime} 2^{\prime} 2} \eta_{1}\right.\\
		&\left.+D_{33^{\prime} 2^{\prime} 23} \eta_{2^{\prime}}-D_{33^{\prime} 2^{\prime} 233^{\prime} \overline{2}} \eta_{3}-D_{33^{\prime} 2^{\prime} 233^{\prime} \bar{2} \bar{2}^{\prime}} \eta_{1^{\prime}}\right] \\
		&-\left[\eta_{1^{\prime}}+D_{2^{\prime}} \eta_{3}+D_{2^{\prime} 2} \eta_{1}+D_{2^{\prime} 23} \eta_{2^{\prime}}-D_{2^{\prime} 233^{\prime} \overline{2}} \eta_{3}\right.\\
		&\left.-D_{2^{\prime} 233^{\prime} \overline{2} \overline{2}^{\prime}} \eta_{1^{\prime}}+D_{2^{\prime} 233^{\prime} \overline{2} \overline{2}^{\prime}} \eta_{1}+D_{2^{\prime} 233^{\prime} \overline{2} \overline{2}^{\prime} 3} \eta_{2^{\prime}}\right].
	\end{aligned}
\end{equation}
As pointed out by some authors, the above combination occupies a shorter space in the time domain, which can effectively avoid the influence of instrument gaps and glitches.
Moreover, in the high-frequency region, its gravitational wave response and noise power spectral density also display better performance, as discussed in~\cite{Geo-TDI-1}.

The coefficients $q_{1}$ and $q_{1^{'}}$ of the above solution can be extracted and found to be
\begin{equation}\label{newM16q}
	\begin{aligned}
		q_{1} &= 1-b+ab-ba\bar{b},\\
		q_{1^{'}} &= - (1-a-ba\bar{b}+aba\bar{b}).
	\end{aligned}
\end{equation}
By substituting the solution into Eq.~\eqref{nMTDI}, the residual is found to be $\Delta=[ba\bar{b}, a]$. 

At first glance, it seems that the form of the residual does not satisfy the condition Eq.~\eqref{permuXY}, as the two terms in the commutator do not even have the same length. 
They are, nonetheless, indeed related by a permutation by making use of Eq.~\eqref{DbarDef} and, therefore
\begin{equation}\label{resM16q}
\Delta = [ba\bar{b}, a]=[ba\bar{b}, a\bar{b}b] .
\end{equation}
Moreover, following the train of thought of the combinatorial approach discussed in the previous section, one can explicitly show how the residual Eq.~\eqref{resM16q} is related to the equation Eq.~\eqref{nMTDI}.
This is achieved by successively subtracting out terms proportional to the factor $(1-a)$ or $(1-b)$ until the last trailing ``$1$'', which gives 
\begin{equation}\label{newMTDI}
	ba\bar{b}a=-ba\bar{b}(1-a)+ba\bar{b}(1-b)-b(1-a)-(1-b)+1,
\end{equation}
and
\begin{equation}
	aba\bar{b}=aba\bar{b}(1-b)-ab(1-a)-a(1-b)-(1-a)+1 .
\end{equation}
Subsequently, one has
\begin{equation}\label{break}
	\Delta=(1-b+ab-ba\bar{b})(1-a)-(1-a-ba\bar{b}+aba\bar{b})(1-b) ,
\end{equation}
which essentially indicates that the coefficients of Eq.~\eqref{nMTDI} are given by Eq.~\eqref{newM16q}.
It is not difficult to observe that the above derivations are generally applicable to solving the TDI equation of the following form
\begin{equation}\label{nMTDIgen}
	\alpha(1-a)+\beta(1-b)=0 ,
\end{equation}
where the forms of $a$ and $b$ are {\it not} specified.
Its valid solution is associated with a commutator of a rather arbitrary form which is composed of two monomials of time-delay and time-advance operators
\begin{equation}\label{DeltaNew}
\Delta = \left[s_{n}, \bar{s}_{n}\right]\equiv t_{2n} - \bar{t}_{2n},
\end{equation}
once it satisfies the condition Eq.~\eqref{permuXY}.
Also, as demonstrated by the above example, the monomials in question do not necessarily have the same counts of $a$s and $b$s.
Naturally, the two terms in the commutator are not obliged to be related by the interchange operation $\mathcal{I}$.

\begin{figure*}
\centering
\includegraphics[width=1\textwidth]{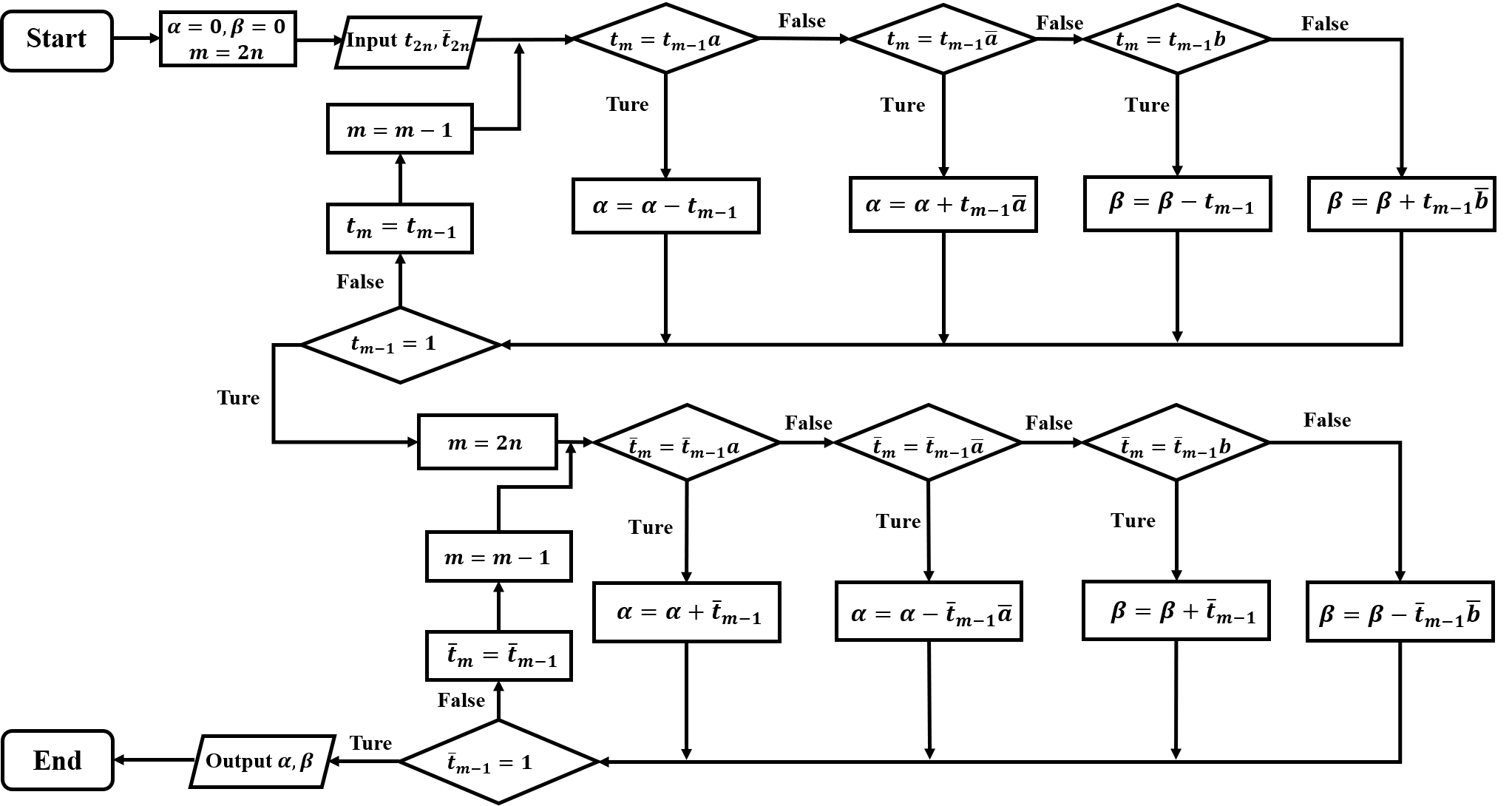}
\caption{\label{process2} 
The control-flow diagram of the generalized algorithm proposed in the present study.}
\end{figure*}

For a given commutator, the extended combinatorial algorithm is summarized as follows, which is also illustrated by the control-flow diagram Fig.~\ref{process2}.
\begin{enumerate}
\item Initiate $\alpha=0$ and $\beta=0$
\item Consider the first term on the r.h.s. of Eq.~\eqref{DeltaNew}, $t_{2n}$, it is a monomial of degree $2n$.
It is noted that $t_{2n}$ ends in either $a$, $\bar{a}$, $b$, or $\bar{b}$, namely, $t_{2n}=t_{2n-1}a$, $t_{2n}=t_{2n-1}\bar{a}$, $t_{2n}=t_{2n-1}b$, or $t_{4n}=t_{2n-1}\bar{b}$. 
\begin{itemize}
\item If $t_{2n}=t_{2n-1}a$, let $\alpha=\alpha-t_{2n-1}$;
\item If $t_{2n}=t_{2n-1}\bar{a}$, let $\alpha=\alpha+t_{2n-1}\bar{a}$;
\item If $t_{2n}=t_{2n-1}b$, let $\beta=\beta-t_{2n-1}$;
\item If $t_{2n}=t_{2n-1}\bar{b}$, let $\beta=\beta+t_{2n-1}\bar{b}$;
\end{itemize}
\item Repeat the above procedure 2 for $t_{2n-1}$ until the degree of the monomial vanishes, namely, $t_{0}$.
We note that $t_{0}=1$.
\item Perform the steps similar to 2 and 3 for $\bar{t}_{2n}$, 
\begin{itemize}
\item If $\bar{t}_{2n}=\bar{t}_{2n-1}a$, let $\alpha=\alpha+\bar{t}_{2n-1}$;
\item If $\bar{t}_{2n}=\bar{t}_{2n-1}\bar{a}$, let $\alpha=\alpha-\bar{t}_{2n-1}\bar{a}$;
\item If $\bar{t}_{2n}=\bar{t}_{2n-1}b$, let $\beta=\beta+\bar{t}_{2n-1}$;
\item If $\bar{t}_{2n}=\bar{t}_{2n-1}\bar{b}$, let $\beta=\beta-\bar{t}_{2n-1}\bar{b}$;
\end{itemize}
until the degree of the monomial vanishes. 
It is noted that $\bar{t}_{0}=1$. 
\item Both $t_{2n}-t_0$ and $\bar{t}_{2n}-\bar{t}_0$ can be rewritten as a summation of multipliers of either $(1-a)$ or $(1-b)$. 
Therefore, the TDI coefficients of Eq.~\eqref{nMTDIgen} are the resulting $\alpha$ and $\beta$.
\end{enumerate}

As an example, for the case of Eq.~\eqref{resM16q}, we have $t_{4}=ba\bar{b}a$ and $\bar{t}_{4}=aba\bar{b}$.
Subsequently, we have $t_{3}=ba\bar{b}$, $t_{2}=ba$, $t_{1}=b$, $t_{0}=1$; $\bar{t}_3=aba$, $\bar{t}_2=ab$, $\bar{t}_1=a$, and $\bar{t}_0=1$. 
So that $\alpha=-t_{3}-t_1+\bar{t}_2+\bar{t}_0=-ba\bar{b}+ab-b+1$ and $\beta=t_2\bar{b}-t_0-\bar{t}_3\bar{b}+\bar{t}_1=-aba\bar{b}+ba\bar{b}+a-1$, consistent with Eq.~\eqref{break}.

\section{Applications to TDI combinations}\label{sec5}

In this section, we explore the applications of the extended combinatorial approach developed in the last section.
We elaborate on the Monitor-type TDI solutions in the main text and relegate the remaining results on other TDI solutions to Appendix~\ref{B}. 
In particular, we present a novel class of Sagnac-inspired combinations, which cannot be straightforwardly obtained using the geometric TDI.
Moreover, in Appendix~\ref{C}, we give explicit forms of the relevant lower-order commutators in an exhaustive fashion that furnish the TDI solution investigated in the present study.
Appendix~\ref{D} is devoted to discussing how to construct specific higher-order TDI solutions based on existing lower-order ones.

\subsection{The Monitor-type combinations}\label{sec5.1}

From the specific form of the modified first-generation and modified second-generation Monitor-E combinations~\cite{TDI2-1}, namely,
\begin{equation}
	\begin{aligned}
		E_{1} (t) &= [(1-D_{311^{'}\bar{3}})\eta_{1}+D_{3}\eta_{2} + D_{31}\eta_{{3}^{'}}]\\
		&-[(1-D_{2^{'}1^{'}1\bar{2}^{'}})\eta_{1^{'}}+D_{2^{'}}\eta_{3^{'}}+D_{2^{'}1^{'}}\eta_{2}],
	\end{aligned}
\end{equation}
and
\begin{equation}
	\begin{aligned}
		E_{2} (t) &= (1-D_{2^{'}1^{'}1\bar{2}^{'}})[(1-D_{311^{'}\bar{3}})\eta_{1}+D_{3}\eta_{2} + D_{31}\eta_{{3}^{'}}]\\
		&-(1-D_{311^{'}\bar{3}})[(1-D_{2^{'}1^{'}1\bar{2}^{'}})\eta_{1^{'}}+D_{2^{'}}\eta_{3^{'}}+D_{2^{'}1^{'}}\eta_{2}] ,
	\end{aligned}
\end{equation}
one observes the following constraint equations
\begin{equation}\label{Moncons}
	q_{3}=0,q_{2^{'}}=0.
\end{equation}

Substituting Eq.~\eqref{Moncons} into Eq.~\eqref{TDIeq}, one finds
\begin{subequations}\label{Monitor}
	\begin{align}
		&q_{1}+q_{1^{'}}=0,\label{aa}\\
		&q_{2}-q_{3^{'}}D_{1^{'}}-q_{1}D_{3}=0,\label{bb}\\
		&q_{3^{'}}-q_{1^{'}}D_{2^{'}}-q_{2}D_{1}=0.\label{cc}
	\end{align}
\end{subequations}
Eliminating Eqs.~\eqref{bb} and \eqref{cc} by substituting the forms of $q_{1}$ and $q_{1'}$ into Eq.~\eqref{aa}, one finds the desired equation
\begin{equation}\label{MonitorTDI}
	q_{2}(D_{\bar{3}}-D_{1\bar{2}^{'}})+q_{3^{'}}(D_{\bar{2}^{'}}-D_{1^{'}\bar{3}})=0.
\end{equation}
The above equation is essentially Eq.~\eqref{nMTDIgen} by recognizing $\alpha=q_{2}D_{\bar{3}}, \beta=q_{3^{'}}D_{\bar{2}^{'}}$, $a=D_{31\bar{2}^{'}}$, and $b=D_{2^{'}1^{'}\bar{3}}$.
To proceed, the algorithm proposed in the Sec.~\ref{sec4.2} can be applied to extract the TDI coefficients of the Monitor-type combination for a given valid commutator consisting of $a$, $b$, $\bar{a}$, $\bar{b}$. 

For instance, for the commutator $[ba,ab]=[D_{2^{'}1^{'}1\bar{2}^{'}},D_{311^{'}\bar{3}}]$, the corresponding TDI solution expressed in terms of the coefficients $q_2, q_{3^{'}}$ reads
\begin{equation}
	\begin{aligned}
		q_{2} &= (1-D_{2^{'}1^{'}\bar{3}}-D_{2^{'}1^{'}1\bar{2}^{'}}+D_{311^{'}\bar{3}2^{'}1^{'}\bar{3}})D_{3},\\
		q_{3^{'}} &= -(1-D_{31\bar{2}^{'}}-D_{311^{'}\bar{3}}+D_{2^{'}1^{'}1\bar{2}^{'}31\bar{2}^{'}})D_{2^{'}} ,
	\end{aligned}
\end{equation}
which is the modified second-generation standard Monitor-E combination.

As a second example, consider $[ba\bar{b},a]\bar{a}$, the corresponding TDI solution is found to be
\begin{equation}\label{aMon}
	\begin{aligned}
 		q_{2}&=\alpha D_{3} =(1-D_{2^{'}1^{'}\bar{3}}+D_{311^{'}\bar{3}}-D_{311'1\bar{2}'3\bar{1}'\bar{1}\bar{3}})D_{3},\\
		q_{3^{'}} &= \beta D_{2^{'}} = -(1-D_{31\bar{2}^{'}}-D_{2^{'}1^{'}1\bar{2}^{'}3\bar{1}^{'}\bar{2}^{'}}+D_{311^{'}1\bar{2}^{'}3\bar{1}^{'}\bar{2}^{'}})D_{2^{'}} .
	\end{aligned}
\end{equation}
Substitute Eqs.~\eqref{aMon} into Eq.~\eqref{aa} and Eq.~\eqref{cc} for the remaining coefficients $q_{1}$ and $q_{1^{'}}$, one finds
\begin{equation}
	\begin{aligned}
		{E}({t})&=\left(1-{D}_{2^{\prime} 1^{\prime} 1 \overline{2}^{\prime}}+{D}_{311^{\prime} 1 \overline{2}^{\prime}}-{D}_{{2}^{\prime} 1^{\prime} 1 \overline{2}^{\prime} 3 \overline{1}^{\prime} \overline{2}^{\prime}}\right) \eta_{1}\\
		&+\left(1-D_{2^{'}1^{'}\bar{3}}+D_{311^{'}\bar{3}}-D_{311'1\bar{2}'3\bar{1}'\bar{1}\bar{3}}\right) D_{3} \eta_{2}\\
		&-\left(1-{D}_{2^{\prime} 1^{\prime} 1 \overline{2}^{\prime}}+{D}_{311^{\prime} 1 \overline{2}^{\prime}}-{D}_{2^{\prime} 1^{\prime} 1 \overline{2}^{\prime} 3 \overline{1}^{\prime} \overline{2}^{\prime}}\right) \eta_{1^{\prime}}\\
		&-\left(1-D_{31\bar{2}^{'}}-D_{2^{'}1^{'}1\bar{2}^{'}3\bar{1}^{'}\bar{2}^{'}}+D_{311^{'}1\bar{2}^{'}3\bar{1}^{'}\bar{2}^{'}}\right) D_{2^{\prime}} \eta_{3^{\prime}} ,
	\end{aligned}
\end{equation}
which is recognized as the alternative Monitor-E combination (c.f. Fig.~9(b) of Ref.~\cite{Geo-sister}).

The same procedure can be carried out for other commutator such as $[ab^2,bab]$, $[a\bar{b},\bar{b}a]$, and $[aab,aba]$, which subsequently gives rise to a class of (higher-order, from the geometric TDI perspective) Monitor-E combinations. 
Moreover, by introducing cyclic permutations between the indices: $3\rightarrow2,2^{'}\rightarrow 1^{'}$ and $3\rightarrow1,2^{'}\rightarrow 3^{'}$, one obtains Monitor-F-type and Monitor-G combinations.
We refer to the above solutions obtained by considering the constraints Eq.~\eqref{Moncons} as Monitor-type solutions.

We point out that this strategy can be utilized to derive other types of TDI combinations, such as the Beacon, Relay, and Sagnac ones.
We relegate the derivations to Appendix~\ref{B} of the paper. 
In Tab.~\ref{t3}, we summarize the relevant types of TDI solutions that can be dealt with using the present approach and the associated constraint equations.
Moreover, in Appendix~\ref{C}, we enumerate the explicit forms of the lower-order commutators, which furnish the TDI solutions.

\begin{table*}
	\centering
	\caption{The TDI combinations and the associated constraints that are adopted systematically by the proposed algorithm.}
	\label{t3}
	\newcommand{\tabincell}[2]{\begin{tabular}{@{}#1@{}}#2\end{tabular}}
	\renewcommand\arraystretch{2}
	\begin{tabular}{ccc}
		\hline
		\hline
		TDI combination&Constraint equation&TDI equation\\ \hline
		Sagnac-$\alpha$ combination& $q_2=q_1 D_3$, $q_3=q_1 D_{31}$ & $q_1 (1-D_{312})+q_{1^{'}}(1-D_{2^{'}1^{'}3^{'}})=0$\\
		Fully symmetric Sagnac combination & $q_{3}=-q_{3^{'}}$, $q_{2}=-q_{2^{'}}$ & $q_{1}(1-D_{3\bar{1}^{'}2})+q_{1^{'}}(1-D_{2^{'}\bar{1}3^{'}})=0$\\
		Michelson-X combination& $q_{2}=0$, $q_{3^{'}}=0$ & $q_{1}(1-D_{33^{'}})+q_{1^{'}}(1-D_{2^{'}2})=0$\\
		Relay-U combination& $q_{2}=0$, $q_{3}=0$ & $q_{2^{'}}(D_{\bar{3}}-D_{3^{'}})+q_{3^{'}}(D_{\bar{2}^{'}}-D_{1^{'}\bar{3}})=0$\\
		Beacon-P combination& $q_{3}=0$, $q_{3^{'}}=0$ & $q_{1}(1-D_{33^{'}})+q_{1^{'}}(1-D_{2^{'}\bar{1}3^{'}})=0$\\
		Monitor-E combination & $q_{3}=0$, $q_{2^{'}}=0$ & $q_{2}(D_{\bar{3}}-D_{1\bar{2}^{'}})+q_{3^{'}}(D_{\bar{2}^{'}}-D_{1^{'}\bar{3}})=0$\\ \hline
	\end{tabular}
\end{table*}

Before closing this subsection, we compared the derived TDI solutions with those obtained using the geometric TDI approach.
In Ref.~\cite{Geo-sister}, a total of nine sixteen-link modified second-generation TDI combinations were reported. 
All these solutions can be retrieved using the present method, and the associated commutator and link trajectory are enumerated in Tab.~\ref{t2}.

	\begin{table*}
	\centering
	\caption{The relevant commutators of the nine sixteen-link modified second-generation TDI combinations.
	The superscript of a TDI combination indicates the number of links, and the subscript represents the index of the combination for a specific solution family~\cite{Geo-sister}.}
	\label{t2}
	\newcommand{\tabincell}[2]{\begin{tabular}{@{}#1@{}}#2\end{tabular}}
	\renewcommand\arraystretch{2}
	\begin{tabular}{ccc}
		\hline
		\hline
		TDI combination&Commutator&Laser link trajectory\\ \hline
		\multirow{2}{*}{$[X]^{16}_{1}$}& $[ba,ab]$ & \multirow{2}{*}{$1 \leftarrow 2 \leftarrow 1 \leftarrow 3 \leftarrow 1 \leftarrow 3 \leftarrow 1 \leftarrow 2 \leftarrow 1 \to 3 \to 1 \to 2 \to 1 \to 2 \to 1 \to 3 \to 1$}\\
		& $a=D_{33^{'}},b=D_{2^{'}2}$ & \\ \cline{2-3}
		\multirow{2}{*}{$[X]^{16}_{2}$} & $[ba\bar{b},a]$ & \multirow{2}{*}{$1 \leftarrow 2 \leftarrow 1 \leftarrow 3\leftarrow 1 \leftarrow 2 \leftarrow 1 \to 3 \to 1 \to 2 \to 1 \leftarrow 3 \leftarrow 1 \to 2 \to 1 \to 3 \to 1$}\\
		& $a=D_{33^{'}},b=D_{2^{'}2}$ & \\ \cline{2-3}
		\multirow{2}{*}{$[U]^{16}_{1}$}& $[ba,ab]$ & \multirow{2}{*}{$1 \leftarrow 2 \leftarrow 1 \leftarrow 3 \leftarrow 2 \to 1 \leftarrow 3 \leftarrow 2 \leftarrow 1 \leftarrow 2 \to 3 \to 1 \to 2 \to 1 \to 2 \to 3 \to 1$}\\
		& $a=D_{33^{'}},b=D_{2^{'}1^{'}\bar{3}}$ & \\ \cline{2-3}
		\multirow{2}{*}{$[U]^{16}_{2}$}& $[ba\bar{b},a]$  & \multirow{2}{*}{$1 \leftarrow 2 \leftarrow1 \leftarrow 3 \leftarrow 2 \leftarrow 1 \leftarrow 2 \to 3 \to 1 \to 2 \to 1 \leftarrow 3 \leftarrow 2 \to 1 \to 2 \to 3 \to 1$}\\
		& $a=D_{33^{'}},b=D_{2^{'}1^{'}\bar{3}}$ & \\ \cline{2-3}
		\multirow{2}{*}{$[U]^{16}_{3}$}& $[ab\bar{a},b]\bar{b}$  & \multirow{2}{*}{$1 \leftarrow 2 \leftarrow 1 \leftarrow 3 \leftarrow 2 \to 1 \to 2 \to 1 \leftarrow 3 \leftarrow 2 \leftarrow 1 \leftarrow 2 \to 3 \to 1 \to 2 \to 3 \to 1$}\\
		& $a=D_{33^{'}},b=D_{2^{'}1^{'}\bar{3}}$ & \\ \cline{2-3}
		\multirow{2}{*}{$[E]^{16}_{1}$}& $[ba,ab]$ & \multirow{2}{*}{$1 \leftarrow 2 \leftarrow 3 \leftarrow 2 \to 1 \leftarrow 3 \leftarrow 2 \leftarrow 3 \to 1 \leftarrow 2 \to 3 \to 2 \to 1 \leftarrow 3 \to 2 \to 3 \to 1$}\\
		& $a=D_{31\bar{2}^{'}},b=D_{2^{'}1^{'}\bar{3}}$ & \\ \cline{2-3}
		\multirow{2}{*}{$[E]^{16}_{2}$} & $[ba\bar{b},a]\bar{a}$  & \multirow{2}{*}{$1 \leftarrow 2 \leftarrow 3 \leftarrow 2 \leftarrow 3 \to 1 \leftarrow 2 \to 3 \to 2 \to 1 \leftarrow 3 \leftarrow 2 \to 1 \leftarrow 3 \to 2 \to 3 \to 1$} \\
		& $a=D_{31\bar{2}^{'}},b=D_{2^{'}1^{'}\bar{3}}$ & \\ \cline{2-3}
		\multirow{2}{*}{$[P]^{16}_{1}$} & $[ba\bar{b},a]$  & \multirow{2}{*}{$1 \leftarrow 2 \leftarrow 1 \leftarrow 3\to 2 \leftarrow 1 \leftarrow 2 \leftarrow 3 \to 1 \to 2 \to 1 \leftarrow 3 \to 2 \to 1 \to 2 \leftarrow 3 \to 1$}\\
		& $a=D_{33^{'}},b=D_{2^{'}\bar{1}3^{'}}$ & \\ \cline{2-3}
		\multirow{2}{*}{$[P]^{16}_{2}$} & $[b,ab\bar{a}]$ & \multirow{2}{*}{$1\leftarrow 2 \leftarrow 1 \leftarrow 3 \to 2 \to 1 \leftarrow 3 \to 2 \leftarrow 1 \leftarrow 2 \leftarrow 3 \to 1 \to 2 \to 1 \to 2 \leftarrow 3 \to 1$}\\
		& $a=D_{33^{'}},b=D_{2^{'}\bar{1}3^{'}}$ & \\ \hline
	\end{tabular}
\end{table*}

\subsection{The fully symmetric Sagnac-type combinations}\label{5.2}

Unlike other, the fully symmetric Sagnac combinations constitute a unique family of solutions.
Their distinctive nature partly resides in the fact that these solutions do not often possess a straightforward geometric TDI interpretation.
In the last two subsections, we focus on these peculiar solutions.
The present subsection will be devoted to retrieving the existing fully symmetric TDI combinations already explored in the literature.
In the following subsection, we elaborate on some novel Sagnac-inspired solutions that have not been reported.

Again, we first discuss the constraint equations through which the appropriate equation Eq.~\eqref{nMTDIgen} can be established.
The available TDI solutions of fully symmetric Sagnac-type are then shown to be consistent with the derived equation.
Subsequently, the algorithm is employed straightforwardly to solve for modified second-generation solutions.

The first-generation fully symmetric Sagnac solution is given by
\begin{equation}\label{1gfS}
	\zeta_{1}({t})={D}_{1} \eta_{1}+D_{2} \eta_{2}+D_{3} \eta_{3}-D_{1^{\prime}} \eta_{1^{\prime}}-D_{2^{\prime}} \eta_{2^{\prime}}-D_{3^{\prime}} \eta_{3^{\prime}},
\end{equation}
for which the residual laser frequency noise reads
\begin{equation}
\Sigma_{i}(D_{i^{'}}-D_{i}+D_{(i-1)(i+1)}-D_{(i+1)^{'}(i-1)^{'}})p_{i}(t) .
\end{equation}
The modified first-generation fully symmetric Sagnac solution possesses the form~\cite{TDI2-1}
\begin{equation}\label{1.5fs}
	\begin{aligned}
		\zeta_{2} (t) &= (D_{2^{'}3^{'}}-D_{1})(D_{3}\eta_{3}-D_{3}\eta_{3^{\prime}}+D_{1^{'}}\eta_{1})\\
		&-(D_{32}-D_{1^{'}})(D_{1}\eta_{1^{'}}-D_{2^{'}}\eta_{2}+D_{2^{'}}\eta_{2^{'}}),
	\end{aligned}
\end{equation}
where the residual laser frequency noise is found to be
\begin{equation}\label{residual1a5}
[D_{2^{'}3^{'}}-D_{1},D_{32}-D_{1^{'}}]p_{1}(t) .
\end{equation} 
It is noted that Eq.~\eqref{residual1a5} is in accordance with the assertion that an arbitrary commutator between different delay operators vanishes, namely, 
\begin{equation}\label{commutatorVanx}
[D_{i(i^{'})}, D_{j(j^{'})}]=0 . 
\end{equation}
It was pointed out in~\cite{TDI2-1} that although the residual Eq.~\eqref{residual1a5} does not vanish regarding the terms $\dot{L}$, the first-order derivative in time, the noise has been suppressed below the desired level.
In other words, for a second- or higher-generation TDI solution, one would demand the more stringent condition imposed by Eq.~\eqref{permuXY}, which ensures an explicit cancelation of the terms $\dot{L}$ in the residual.  
In what follows, we will first retrieve Eq.~\eqref{1.5fs} using the combinatorial algebraic approach, then explore further by aiming at the modified second-generation results.

For the first-generation TDI, besides Eq.~\eqref{commutatorVanx}, one has
\begin{equation}\label{commutatorVanx1a0}
D_{i^{'}} = D_{i} .
\end{equation}
Subsequently, we consider the following constraints $q_{3}+q_{3^{'}}=0, q_{2}+q_{2^{'}}=0$. 
Note that it is also viable to use the constraint involving $q_{1}$ and $q_{1^{'}}$ by permutating the indices in a cyclic order.
As pointed out in \cite{TDI2-1}, the Sagnac effect from the rotation of the array breaks the uniqueness of the ``$\zeta$-like'' combinations, which incorporate with the constraints $q_{1}+q_{1^{'}}=0$.
By substituting
\begin{equation}\label{fscons}
	q_{3}=-q_{3^{'}}, q_{2}=-q_{2^{'}}
\end{equation}
into Eq.~\eqref{TDIeq}, we have
\begin{equation}\label{TDIeqfS}
	q_{1}(1-D_{3\bar{1}^{'}2})+q_{1^{'}}(1-D_{2^{'}\bar{1}3^{'}})=0 ,
\end{equation}
which is equivalent to Eq.~\eqref{nMTDIgen} by identifying $a=D_{3\bar{1}^{'}2}, b=D_{2^{'}\bar{1}3^{'}}$.
From Eq.~\eqref{1gfS}, one has $q_1=D_{1}$ and $q_{1'}=D_{1^{'}}$, which manifestly satisfies Eq.~\eqref{TDIeqfS} while taking into account Eq.~\eqref{commutatorVanx1a0}.

For the case of modified first-generation TDI, we replace $D_{\bar{1}^{'}}$ and $D_{2^{'}}$ with $D_{3}$ and $D_{\bar{1}}$ in the above expression to obtain
\begin{equation}
	q_{1}(1-D_{\bar{1}^{'}32})+q_{1^{'}}(1-D_{\bar{1}2^{'}3^{'}})=0 .
\end{equation}
Again, it is straightforward to show that $q_1=(D_{2^{'}3^{'}}-D_{1})D_{1^{'}}$ and $q_{1'}=-(D_{32}-D_{1^{'}})D_{1}$ read off from Eq.~\eqref{1.5fs} satisfies the above equation with Eq.~\eqref{commutatorVanx}.

We now use the algorithm proposed in Sec.~\ref{sec4.2} to solve for the modified second-generation fully symmetric Sagnac combinations. 
To start with, consider the commutator $[ba,ab]=[D_{2^{'}\bar{1}3^{'}3\bar{1}^{'}2},D_{3\bar{1}^{'}22^{'}\bar{1}3^{'}}]$, the TDI coefficients are found to be
\begin{equation}
	\begin{aligned}
		q_{1} &= 1-D_{2^{'}\bar{1}3^{'}}-D_{2^{'}\bar{1}3^{'}3\bar{1}^{'}2}+D_{3\bar{1}^{'}22^{'}\bar{1}3^{'}2^{'}\bar{1}3^{'}},\\
		q_{1^{'}}&=-(1-D_{3\bar{1}^{'}2}-D_{3\bar{1}^{'}22^{'}\bar{1}3^{'}}+D_{2^{'}\bar{1}3^{'}3\bar{1}^{'}23\bar{1}^{'}2}).
	\end{aligned}
\end{equation}
The corresponding TDI solution reads
\begin{equation}\label{fs2.5}
	\begin{aligned}
		\zeta(t)&=\left(1-D_{2^{\prime} \overline{1} 3^{\prime}}-D_{2^{\prime} \overline{1} 3^{\prime} 3 \overline{1}^{\prime} 2}+D_{3 \overline{1}^{\prime} 22^{\prime} \overline{1} 3^{\prime} 2^{\prime} \overline{1} 3^{\prime}}\right) \eta_{1}\\
		&+\left(1-D_{3 \overline{1}^{\prime} 2}-D_{3 \overline{1}^{\prime} 22^{\prime} \overline{1} 3^{\prime}}+D_{2^{\prime} \overline{1} 3^{\prime} 3\overline{1}^{\prime} 23 \overline{1}^{\prime} 2}\right) D_{2^{\prime} \overline{1}} \eta_{2}\\
		&+\left(1-D_{2^{\prime} \overline{1} 3^{\prime}}-D_{2^{\prime} \overline{1} 3^{\prime} 3 \overline{1}^{\prime} 2}+D_{3 \overline{1}^{\prime} 22^{\prime} \overline{1} 3^{\prime} 2^{\prime} \overline{1} 3^{\prime}}\right) D_{3 \overline{1}^{\prime}} \eta_{3}\\
		&-\left(1-D_{3\bar{1}^{'}2}-D_{3\bar{1}^{'}22^{'}\bar{1}3^{'}}+D_{2^{'}\bar{1}3^{'}3\bar{1}^{'}23\bar{1}^{'}2}\right) \eta_{1^{\prime}}\\
		&-\left(1-D_{3 \overline{1}^{\prime} 2}-D_{3 \overline{1}^{\prime} 22^{\prime} \overline{1} 3^{\prime}}+D_{2^{\prime} \overline{1} 3^{\prime} 3 \overline{1}^{\prime} 23 \overline{1}^{\prime} 2}\right) D_{2^{\prime} \overline{1}} \eta_{2^{\prime}}\\
		&-\left(1-D_{2^{\prime} \overline{1} 3^{\prime}}-D_{2^{\prime} \overline{1} 3^{\prime} 3 \overline{1}^{\prime} 2}+D_{3 \overline{1}^{\prime} 22^{\prime} \overline{1} 3^{\prime} 2^{\prime} \overline{1} 3^{\prime}}\right) D_{3 \overline{1}^{\prime}} \eta_{3^{\prime}}.
	\end{aligned}
\end{equation}
The above procedure can be similarly applied to other commutators to give rise to other combinations. 
We refer to these combinations associated with the constraints Eq.~\eqref{fscons} as the fully symmetric Sagnac-type solution. 

\begin{figure}[!t]
	\includegraphics[width=0.50\textwidth]{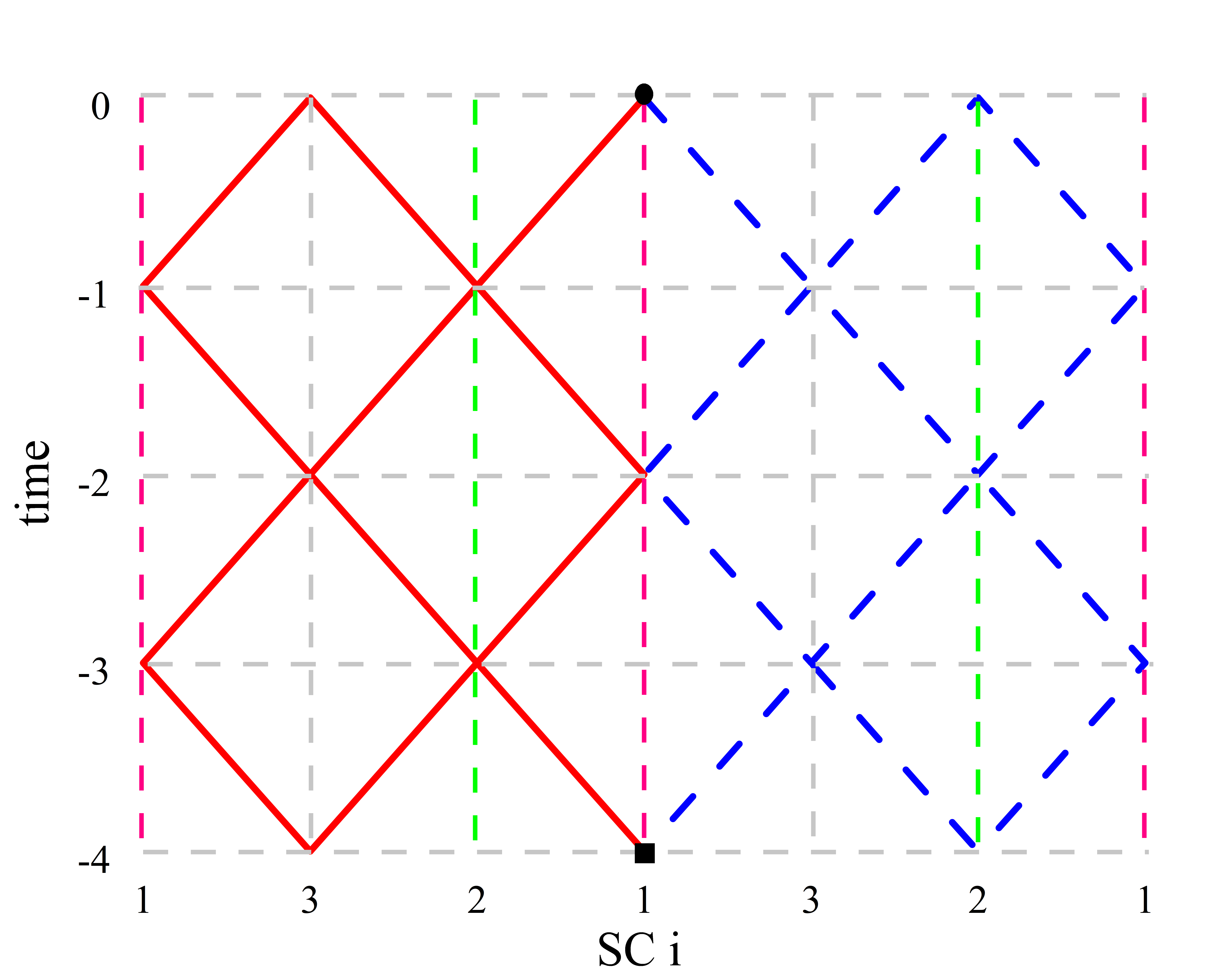}
	\caption{\label{fig2} The space-time diagram for the modified second-generation fully symmetric Sagnac combination given by Eq.~\eqref{fs2.5}.
	The notation is consistent with Ref.~\cite{Geo-sister}.
	The red solid and blue dashed lines represent the two synthetic paths, and the black square and dot indicate the start and end points of the light path, respectively.}
\end{figure}

It is rather interesting to point out that some fully symmetric Sagnac combinations possess a geometric TDI interpretation.
As shown in Fig.~\ref{fig2}, Eq.~\eqref{fs2.5} corresponds to the following space-time diagram regarding the geometric TDI.

\subsection{Novel Sagnac-inspired combinations}\label{5.3}

In this subsection, we elaborate on a novel class of Sagnac-inspired solutions that do not associate with any geometric TDI trajectory.
The distinction is based on the fact that any valid geometric TDI solution demands successive transmissions of laser signal between the spacecraft.
In other words, only specific time-delay or time-advance operators can be applied to a signal emitted from a given optical bench.
For instance, the science data stream $\eta_{1}$ cannot be applied by the time-delay operation associated with the opposite armlength, $D_{1}$.
On the other hand, it is apparent that such a physical constraint is not necessary from an algebraic perspective. 

To this end, we consider the following two constraints $q_{1}=q_{2}D_{1}, q_{2}=q_{3}D_{2}$.
One observes the similarity between the above constraints and those of the  Sagnac-type combinations.
Therefore, the resulting solutions will be referred to as Sagnac-inspired.
Subsequently, Eq.~\eqref{TDIeq} gives
\begin{subequations}
	\begin{align}
		q_{3} D_{21}+q_{1^{\prime}}-q_{2^{\prime}} {D}_{3^{\prime}}-q_{3} {D}_{2}&=0, \label{aaa}\\
		q_{3} D_{2}+q_{2^{\prime}}-q_{3^{\prime}} {D}_{1^{\prime}}-q_{3} D_{213}&=0, \label{bbb}\\
		q_{3}+q_{3^{\prime}}-q_{1^{\prime}} {D}_{2^{\prime}}-q_{3} D_{21}&=0 .\label{ccc}
	\end{align}
\end{subequations}
Eliminating Eqs.~\eqref{aaa} and \eqref{bbb} by substituting the forms of $q_{1'}$ and $q_{2'}$ into Eq.~\eqref{ccc}, one finds the desired equation
\begin{equation}\label{spTDI}
			q_{3}\left(1-A\right)+q_{3^{\prime}}\left(1-b\right)=0,
\end{equation}
where $A= D_{2133^{\prime} 2^{\prime}}-D_{23^{\prime} 2^{\prime}}+D_{22^{\prime}}-D_{212^{\prime}}+D_{21}=\sum_i a_i$ and $b={D}_{1^{\prime} 3^{\prime} 2^{\prime}}$. 
Since $A$ is not a monomial but a polynomial, we note that the condition Eq.~\eqref{permuXY} is not satisfied.
However, one may argue that the vanishing condition of the associated commutator can be generalized to include polynomial, such as $[bA, Ab]$, by noticing
\begin{equation}\label{spTDI1}
			[bA, Ab]=\left[b\left(\sum_i a_i\right), \left(\sum_j a_j\right)b\right]=\sum_{ij}[b a_i, a_j b]\sim \sum_{i}[b a_i, a_i b] ,
\end{equation}
where ``$\sim$'' means identical as an operator when applied to an arbitrary function of time.
This is because the cross terms vanish
\begin{equation}\label{spTDI}
			[b a_i, a_j b]+[b a_j, a_i b]\sim 0,
\end{equation}
for $i\ne j$.
To be specific, one has
\begin{equation}\label{spTDI}
			\left\{[D_b D_{a_i}, D_{a_j} D_b]+[D_b D_{a_j}, D_{a_i} D_b]\right\}\phi(t)=0 ,
\end{equation}
whose validity will be shown shortly.
The r.h.s. of Eq.~\eqref{spTDI1} subsequently vanishes as it satisfies the condition Eq.~\eqref{permuXY}.

The above result is a special case of a generalized relation of Eq.~\eqref{sys2}, which applies to polynomials $A=\sum_i^n a_i$ and $B=\sum_j^m b_j$:
\begin{equation}\label{sys3}
	\left[AB, BA\right]\phi(t)=\left[\sum_{i=1}^{n} a_i \sum_{j=1}^{m} b_j, \sum_{k=1}^{m} b_k \sum_{l=1}^{n} a_l\right] \phi(t)
	=\sum_{ij}\left[a_i b_j, b_j a_i\right]\phi(t) ,
\end{equation}
to the first-order time-derivative terms $\dot{L}$.
To confirm that all the cross terms vanish, we show that they cancel out in pairs.
To be specific, there are $mn(mn-1)/2$ pairs of the form
\begin{equation}
	\Big([a_{i}b_{j},b_{k}a_{l}]+[a_{l}b_{k},b_{j}a_{i}]\Big)\phi(t) ,
\end{equation}
where either $i\ne l$ or $j\ne k$.
Using Eq.~\eqref{sys2}, we have
\begin{equation}
	\begin{aligned}
		&{\left[a_{i} b_{j}, b_{k} a_{l}\right] \phi(} t) \\
		&=\left[\left(L_{a_{i}}+L_{b_{j}}\right) \left(\dot{L}_{b_{k}}+\dot{L}_{a_{l}}\right)\right.\\
		&\left.-\left(L_{b_{k}}+L_{a_{l}}\right) \left(\dot{L}_{a_{i}}+\dot{L}_{b_{j}}\right)\right]\\ &\dot{\phi}\left(t-\left(L_{a_{i}}+L_{b_{j}}\right)\right.
		\left.-\left(L_{b_{k}}+L_{a_{l}}\right)\right),
	\end{aligned}
\end{equation}
and similarly,
\begin{equation}
	\begin{aligned}
		&{\left[a_{l} b_{k}, b_{j} a_{i}\right] \phi(} t) \\
&=\left[\left(L_{a_{l}}+L_{b_{k}}\right) \left(\dot{L}_{b_{j}}+\dot{L}_{a_{i}}\right)\right.\\
&\left.-\left(L_{b_{j}}+L_{a_{i}}\right) \left(\dot{L}_{a_{l}}+\dot{L}_{b_{k}}\right)\right]\\ &\dot{\phi}\left(t-\left(L_{a_{l}}+L_{b_{k}}\right)\right.
\left.-\left(L_{b_{j}}+L_{a_{i}}\right)\right),
	\end{aligned}
\end{equation}
where $L_{a_{i}},L_{b_{j}},\dot{L}_{a_{i}},\dot{L}_{b_{j}}$ represent the armlengths and their rates of change regarding the associated time-translation operations. 
The sum of the two terms manifestly vanishes.

Based on the above arguments, one applies the commutator and finds
\begin{equation}
	\begin{aligned}
		q_{3}&=1-b-bA+Ab^{2},\\
		q_{3^{'}}&=-(1-A-Ab+bA^{2}).
	\end{aligned}
\end{equation}
The remaining coefficients are found to be
\begin{equation}
	\begin{aligned}
		q_{2^{\prime}}&=-\left(1-A-A b+b A^{2}\right) {D}_{1^{\prime}}\\
		&+\left(1-b-b A+A b^{2}\right) D_{2}\left(D_{13}-1\right), \\
		q_{1^{\prime}}&=(-(1-A-A b+b A^{2}) {D}_{1^{\prime}}\\
		&+(1-b-b A+A b^{2}) D_{2}(D_{13}-1)) {D}_{3^{\prime}}\\
		&+(1\left.-b-b A+A b^{2}\right) {D}_{2}\left(1-D_{1}\right), \\
		{q}_{2}&=\left(1-b-b A+A b^{2}\right) D_{2}, \\
		{q}_{1}&=\left(1-b-b A+A b^{2}\right) D_{21}.
	\end{aligned}
\end{equation}
Due to the terms such as $D_{21}\eta_{1}$ in $q_{1}\eta_{1}$, $D_{2}\eta_{2}$ in $q_2\eta_{2}$, the above solution cannot be straightforwardly addressed by the geometric TDI approach. 

\begin{figure}[!t]
	\includegraphics[width=0.5\textwidth]{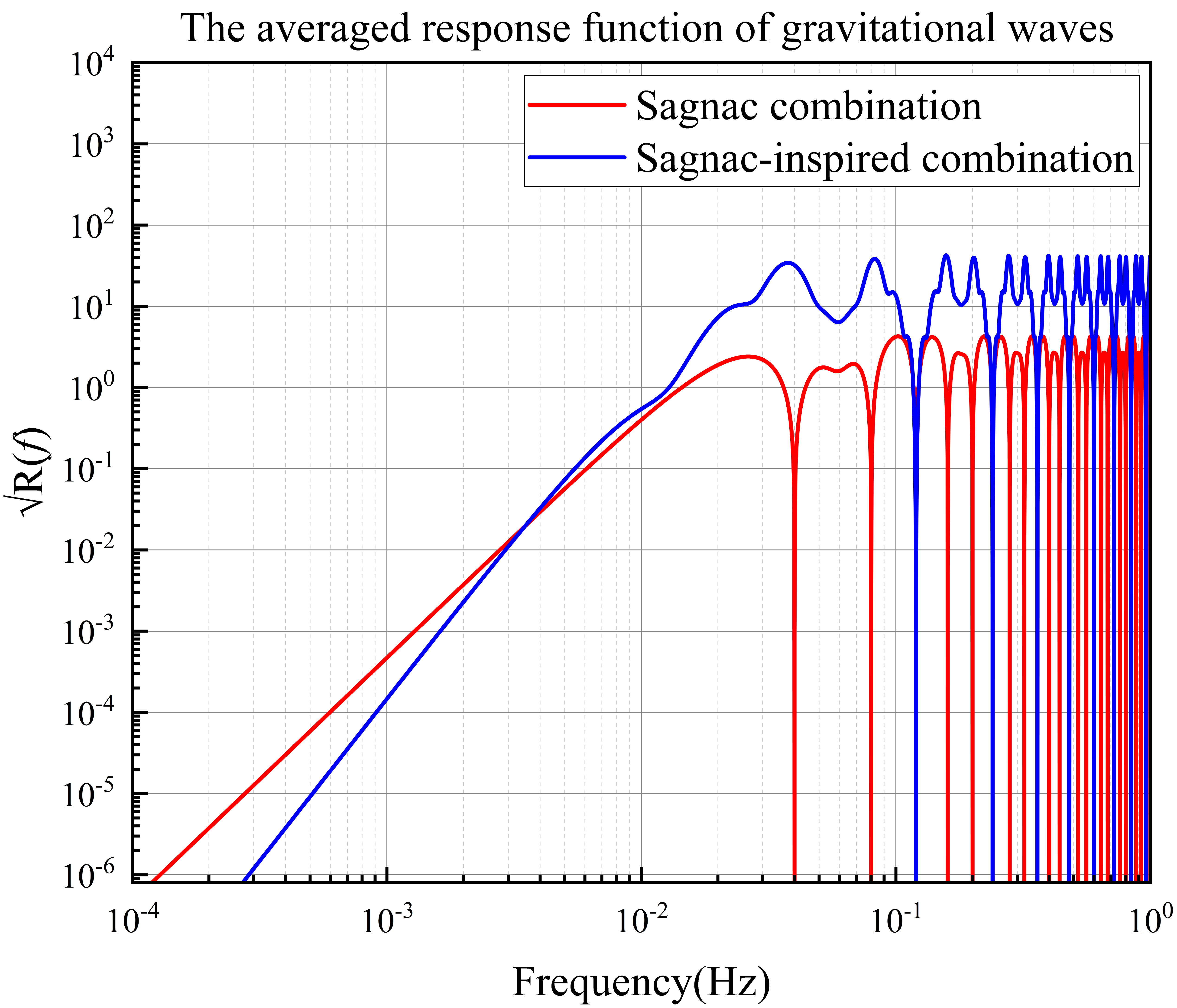}
	\caption{\label{fig3}The gravitational waves averaged response functions of the Sagnac and Sagnac-inspired combinations.}
\end{figure}

\begin{figure}[!t]
	\includegraphics[width=0.5\textwidth]{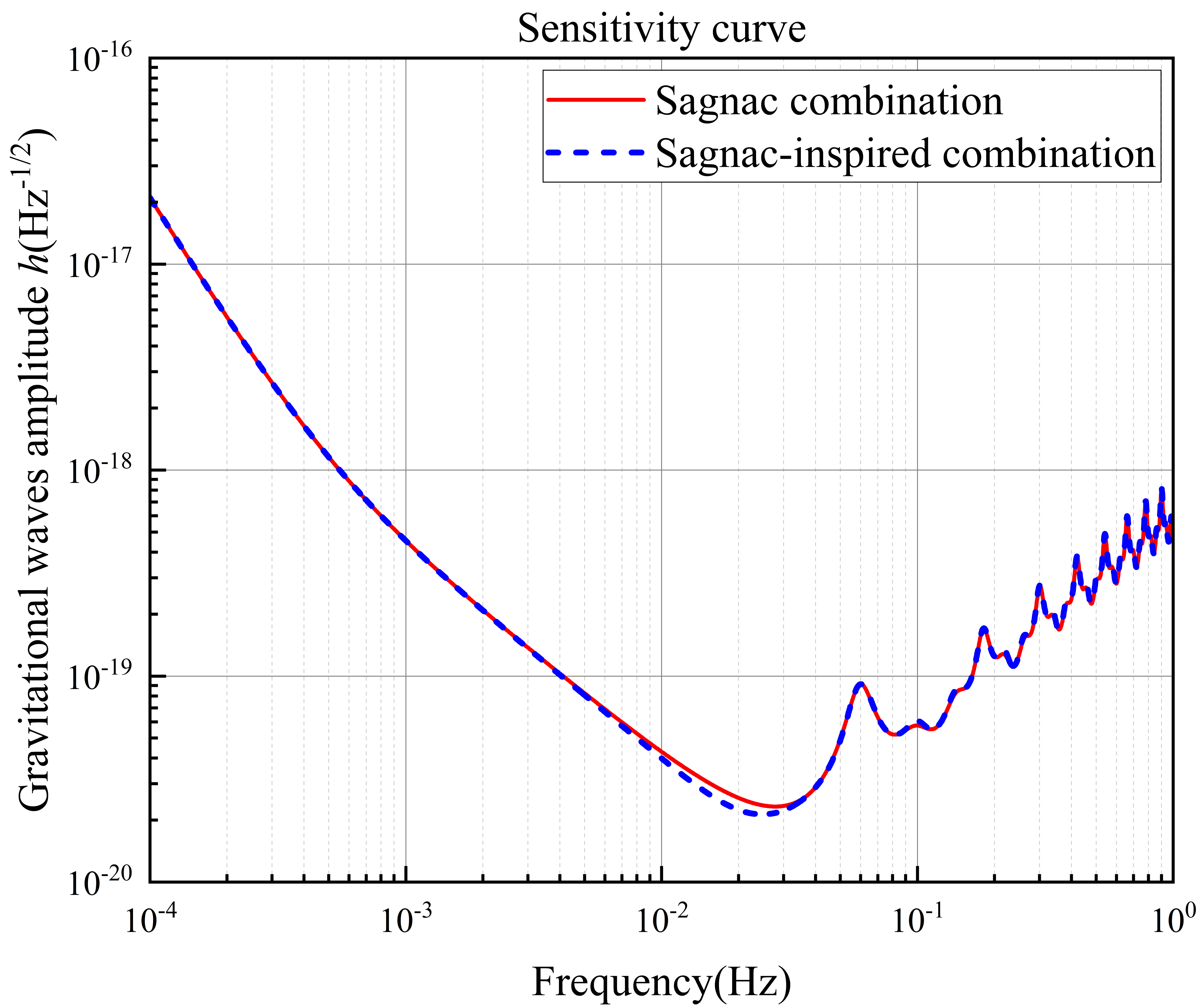}
	\caption{\label{fig4}Sensitivity curves of the Sagnac and Sagnac-inspired combinations.}
\end{figure}

To analyze the performance of the combination obtained by the end of the last section, we evaluate its response function and sensitivity curve.
They are shown in Fig.~\ref{fig3} and~\ref{fig4} in comparison with those of the conventional Sagnac combinations derived in~\cite{sens, sens2}.
Also, we adopt typical parameters in LISA mission, by assuming that armlength $L=2.5\times 10^{6}\mathrm{km}$ and the corresponding ASDs of the test mass and shot noise: $s^{\mathrm{LISA}}_{a}=3\times 10^{-15}\mathrm{m}\cdot \mathrm{s^{-2}}/\sqrt{\mathrm{Hz}}$ and $s^{\mathrm{LISA}}_{x}=15\times 10^{-12}\mathrm{m}/\sqrt{\mathrm{Hz}}$ \cite{Robson_2019}. 
The specific forms of the averaged response functions of gravitational waves and noise power spectral densities are given in Appendix~\ref{E}.
For the averaged response function, the Sagnac-inspired combination shows some advantage in higher frequencies.
On the other hand, the difference in the resultant sensitivity curves between the two combinations is not significant.

It is worth noting, besides the two constraints considered in this section, other choices of constraint conditions such as $q_{1}=q_{2}D_{1}$ and $q_{1^{'}}=q_{2^{'}}D_{1^{'}}$ also leads to solutions without any geometric TDI interpretation.
Therefore, the extended algebraic, combinatorial approach proposed in this study indicates a much broader solution space beyond the scope of the geometric TDI. 


\section{Concluding remarks}\label{sec6}

In this paper, an extended algorithm was proposed to solve for the modified second-generation TDI combinations.
The approach is based on a combinatorial algebraic scheme first introduced by Dhurandhar {\it et al.} applied to the Michelson-type second-generation TDI solutions. 
By introducing the time-advance operator and adopting different constraint conditions, we showed that various second-generation TDI combinations, such as the Monitor, Beacon, and Relay combinations, can be retrieved. 
We also reproduced the fully symmetric Sagnac combinations whose laser-link trajectory cannot be understood in the framework of the geometric TDI algorithm.
Inspired by such constraint conditions, we deduced a novel class of Sagnac-inspired TDI solutions.
On the mathematical side, we generalized the commutation relation and its vanishing condition to a more general context by considering the time-advance operators and their polynomials.
Moreover, we demonstrated that the present scheme is rather general, and the solution space is broader than what has already been explored in the literature.
Subsequently, the present study constitutes an attempt to enrich our understanding of the TDI solutions from an algebraic perspective.

The proposed enumerating scheme is more efficient than geometric TDI, as the latter is potentially time-consuming at higher orders.
Nonetheless, the exhaustive nature of the present algorithm is not entirely clear.
The constraint conditions explicitly delimit the solution.
As a result, it only addresses a specific subspace of the entire solution space.
Also, the commutator relation is only valid up to the first-order time-derivative terms. 
Therefore further generalization to higher-order scenarios is not straightforward.
We plan to address these topics in further studies.

\section*{Acknowledgements}
This work is supported by the National Natural Science Foundation of China (Grant No.11925503), the Postdoctoral Science Foundation of China (Grant No.2022M711259), 
Guangdong Major project of Basic and Applied Basic Research (Grant No.2019B030302001), 
Key Laboratory of TianQin Project(Sun Yat-sen University), Ministry of Education,
and the Fundamental Research Funds for the Central Universities, HUST: 2172019kfyRCPY029.
We also gratefully acknowledge the financial support from Brazilian agencies 
Funda\c{c}\~ao de Amparo \`a Pesquisa do Estado de S\~ao Paulo (FAPESP), 
Funda\c{c}\~ao de Amparo \`a Pesquisa do Estado do Rio de Janeiro (FAPERJ), 
Conselho Nacional de Desenvolvimento Cient\'{\i}fico e Tecnol\'ogico (CNPq), 
and Coordena\c{c}\~ao de Aperfei\c{c}oamento de Pessoal de N\'ivel Superior (CAPES).

\appendix

\section{A proof of the commutation relation}\label{A}

In this Appendix, we give proof of Eq.~\eqref{sys1}, which furnishes the basis for its generalization in Sec.~\ref{sec4}. 

Using {\it mathematical induction}, we first show the following {\it proposition} for all nature number $n$: to the first order time-derivative, one has
\begin{equation}\label{mi}
\begin{split}
&P(n): \\
&D_{x_{n}}...D_{x_{1}}\phi(t)=\phi\left(t-\sum^{n}_{i=1}L_{x_{i}}\right)+\dot{\phi}\left(t-\sum^{n}_{i=1}L_{x_{i}}\right)\left(\sum^{n-1}_{j=1}\dot{L}_{x_{j}}\left(\sum^{n}_{k=j+1}L_{x_{k}}\right)\right)-\dot{\phi}\left(t-\sum^{n}_{i=1}L_{x_{i}}\right)t\left(\sum^{n}_{i=1}\dot{L}_{x_{i}}\right).
\end{split}
\end{equation}

Proof: 

It is straightforward to show that $P(1)$ is true
\begin{equation}
\begin{split}
&P(1):\\
&D_{x_{1}}\phi(t)=\phi(t-L_{x_{1}})-\dot{\phi}(t-L_{x_{1}})t\dot{L}_{x_{1}}
\end{split}
\end{equation}

Induction step: if $P(n)$ holds, we have,
\begin{equation}
\begin{aligned}
P(n+1):\\  
  D_{x_{n+1}}D_{x_{n}}...D_{x_{1}}\phi(t)&=D_{x_{n+1}}\left(\phi\left(t-\sum^{n}_{i=1}L_{x_{i}}\right)+\dot{\phi}\left(t-\sum^{n}_{i=1}L_{x_{i}}\right)\left(\sum^{n-1}_{j=1}\dot{L}_{x_{j}}\left(\sum^{n}_{k=j+1}L_{x_{k}}\right)\right)-\dot{\phi}\left(t-\sum^{n}_{i=1}L_{x_{i}}\right)t\left(\sum^{n}_{i=1}\dot{L}_{x_{i}}\right)\right)\\
    &=\phi\left(t-\sum^{n+1}_{i=1}L_{x_{i}}\right)+\dot{\phi}\left(t-\sum^{n+1}_{i=1}L_{x_{i}}\right)\left(\sum^{n}_{j=1}\dot{L}_{x_{j}}\left(\sum^{n+1}_{k=j+1}L_{x_{k}}\right)\right)-\dot{\phi}\left(t-\sum^{n+1}_{i=1}L_{x_{i}}\right)t\left(\sum^{n+1}_{i=1}\dot{L}_{x_{i}}\right),
\end{aligned}
\end{equation}
so that $P(n+1)$ also holds true.
Q.E.D. 

It is readily shown that Eq.~\eqref{mi} implies
\begin{equation}\label{yx}
\begin{aligned}
     D_{y_{n}}...D_{y_{1}}D_{x_{n}}...D_{x_{1}}\phi(t)&=\phi\left(t-\sum^{n}_{i=1}L_{x_{i}}-\sum^{n}_{i=1}L_{y_{i}}\right)-\dot{\phi}\left(t-\sum^{n}_{i=1}L_{x_{i}}-\sum^{n}_{i=1}L_{y_{i}}\right)t\left(\sum^{n}_{i=1}\dot{L}_{x_{i}}+\sum^{n}_{i=1}\dot{L}_{y_{i}}\right)\\
     &+\dot{\phi}\left(t-\sum^{n}_{i=1}L_{x_{i}}-\sum^{n}_{i=1}L_{y_{i}}\right)\left(\sum^{n-1}_{j=1}\dot{L}_{y_{j}}\left(\sum^{n}_{k=j+1}L_{y_{k}}\right)+\sum^{n}_{j=1}\dot{L}_{x_{j}}\left(\sum^{n}_{k=j+1}L_{x_{k}}+\sum^{n}_{k=1}L_{y_{k}}\right)\right),
\end{aligned}
\end{equation}
and
\begin{equation}\label{xy}
\begin{aligned}
     D_{x_{n}}...D_{x_{1}}D_{y_{n}}...D_{y_{1}}\phi(t)&=\phi\left(t-\sum^{n}_{i=1}L_{y_{i}}-\sum^{n}_{i=1}L_{x_{i}}\right)-\dot{\phi}\left(t-\sum^{n}_{i=1}L_{y_{i}}-\sum^{n}_{i=1}L_{x_{i}}\right)t\left(\sum^{n}_{i=1}\dot{L}_{y_{i}}+\sum^{n}_{i=1}\dot{L}_{x_{i}}\right)\\
     &+\dot{\phi}\left(t-\sum^{n}_{i=1}L_{y_{i}}-\sum^{n}_{i=1}L_{x_{i}}\right)\left(\sum^{n-1}_{j=1}\dot{L}_{x_{j}}\left(\sum^{n}_{k=j+1}L_{x_{k}}\right)+\sum^{n}_{j=1}\dot{L}_{y_{j}}\left(\sum^{n}_{k=j+1}L_{y_{k}}+\sum^{n}_{k=1}L_{x_{k}}\right)\right).
\end{aligned}
\end{equation}
By subtracting Eq.~\eqref{yx} from Eq.~\eqref{xy}, one finds Eq.~\eqref{sys1} given in the main text.

\section{The derivations of other TDI combinations}\label{B}

Here, we present the derivations of other TDI combinations using the combinatorial algorithm proposed in this paper.
As the procedure is mainly reminiscent of what is discussed in Sec.~\ref{sec5}, only the essential arguments are given below.

\subsection{Sagnac-type combinations}

From the specific forms of the first-generation Sagnac-$\alpha$ combination
\begin{equation}
	\alpha_{1}(t)=\eta_{1}-\eta_{1^{\prime}}+{D}_{3} \eta_{2}-{D}_{2^{\prime} 1} \eta_{2^{\prime}}+{D}_{31} \eta_{3}-{D}_{2^{\prime}} \eta_{3^{\prime}},
\end{equation}
and the modified second-generation standard Sagnac-$\alpha$ combination
\begin{equation}
	\begin{aligned}
		\alpha_{2}(t) &=\left(D_{2^{\prime} 1^{\prime} 3^{\prime}}-1\right)\left(\eta_{1}+D_{3} \eta_{2}+D_{31} \eta_{3}\right) \\
		&-\left(D_{312}-1\right)\left(\eta_{1^{\prime}}+D_{2^{\prime}} \eta_{3^{\prime}}+D_{2^{\prime} 1^{\prime}} \eta_{2^{\prime}}\right) ,
	\end{aligned}
\end{equation}
it is observed that the coefficients satisfy the following relations
\begin{equation}\label{sagnac}
	{q}_{2}={q}_{1} D_{3}, {q}_{3}=q_{1} D_{31}, {q}_{3^{\prime}}=q_{1^{\prime}} D_{2^{\prime}}, {q}_{2^{\prime}}=q_{1^{\prime}} D_{2^{\prime} 1^{\prime}}.
\end{equation}

The above relations introduce four constraints.
However, it can be shown that these four constraints are not entirely independent from Eq.~\eqref{TDIeq}.
In particular, the second and third lines of Eq.~\eqref{TDIeq} are manifestly linear combinations of the above constraints, and therefore, Eq.~\eqref{sagnac} only provides two independent conditions.
Subsequently, the relevant equation is obtained from the first line of Eq.~\eqref{TDIeq} by substituing the above conditions and eliminating $q_2, q_3, q_{2'}$, and $q_{3'}$.
One finds
\begin{equation}
	q_{1}(1-D_{312})+q_{1^{'}}(1-D_{2^{'}1^{'}3^{'}})=0.
\end{equation}
Again, the above equation is nothing but Eq.~\eqref{nMTDIgen} by identifying $a=D_{312}$ and $b=D_{2^{'}1^{'}3^{'}}$. 
Subsequently, the algorithm may proceed as described in Sec.~\ref{sec3.2}, as long as no time-advance operator is involved.
In this case, one may essentially follow~\cite{D2010} to systematically enumerate the relevant commutators related to the Sagnac-$\alpha$ type combinations.
The simplest choice is $\Delta=[ba,ab]=[D_{2^{'}1^{'}3^{'}312},D_{3122^{'}1^{'}3^{'}}]$, whose the coefficients $q_1,q_{1^{'}}$,
\begin{equation}
	\begin{aligned}
		&q_{1}=1-D_{2^{\prime} 1^{\prime} 3^{\prime}}-D_{2^{\prime} 1^{\prime} 3^{\prime} 312}+D_{3122^{\prime} 1^{\prime} 3^{\prime} 2^{\prime} 1^{\prime} 3^{\prime}}, \\
		&q_{1^{\prime}}=-\left(1-D_{312}-D_{3122^{\prime} 1^{\prime} 3^{\prime}}+D_{2^{\prime} 1^{\prime} 3^{\prime} 312312}\right) ,
	\end{aligned}
\end{equation}
are recognized as those of the modified second-generation Sagnac-$\alpha$ combination given above.

On the other hand, one may consider commutator that contains the time-advance operator.
As an example, for the case of $\Delta=[ba\bar{b},a]$, we arrive at an alternative Sagnac combination
\begin{equation}
	\begin{aligned}
		&\alpha({t})=\left(1-D_{2^{\prime} 1^{\prime} 3^{\prime}}+{D}_{3122^{\prime} 1^{\prime} 3^{\prime}}-D_{2^{\prime} 1^{\prime} 3^{\prime} 312 \overline{3}^{\prime} \overline{1}^{\prime} \overline{2}^{\prime}}\right) \eta_{1}\\
		&+\left(1-D_{2^{\prime} 1^{\prime} 3^{\prime}}+{D}_{3122^{\prime} 1^{\prime} 3^{\prime}}-D_{2^{\prime} 1^{\prime} 3^{\prime} 312 \overline{3}^{\prime} \overline{1}^{\prime} \overline{2}^{\prime}}\right) D_{3} \eta_{2}\\
		&+\left(1-D_{2^{\prime} 1^{\prime} 3^{\prime}}+{D}_{3122^{\prime} 1^{\prime} 3^{\prime}}-D_{2^{\prime} 1^{\prime} 3^{\prime} 312 \overline{3}^{\prime} \overline{1}^{\prime} \overline{2}^{\prime}}\right) D_{31} \eta_{3}\\
		&-\left(1-{D}_{312}-D_{2^{\prime} 1^{\prime} 3^{\prime} 312 \overline{3}^{\prime} \overline{1}^{\prime} \overline{2}^{\prime}}+{D}_{3122^{\prime} 1^{\prime} 3^{\prime} 312 \overline{3}^{\prime} \overline{1}^{\prime} \overline{2}^{\prime}}\right) \eta_{1^{\prime}}\\
		&-\left(1-{D}_{312}-D_{2^{\prime} 1^{\prime} 3^{\prime} 312 \overline{3}^{\prime} \overline{1}^{\prime} \overline{2}^{\prime}}+{D}_{3122^{\prime} 1^{\prime} 3^{\prime} 312 \overline{3}^{\prime} \overline{1}^{\prime} \overline{2}^{\prime}}\right) D_{2^{\prime}} \eta_{2^{\prime}}\\
		&-\left(1-{D}_{312}-D_{2^{\prime} 1^{\prime} 3^{\prime} 312 \overline{3}^{\prime} \overline{1}^{\prime} \overline{2}^{\prime}}+{D}_{3122^{\prime} 1^{\prime} 3^{\prime} 312 \overline{3}^{\prime} \overline{1}^{1} \overline{2}^{\prime}}\right) D_{2^{\prime} 1^{\prime}} \eta_{3^{\prime}}.
	\end{aligned}
\end{equation}

It is readily verified that the commutators $[ab^2,bab]$, $[a\bar{b},\bar{b}a]$, $[aab,aba]$, among others, give rise to a class of higher-order (from the perspective of the geometric TDI) Sagnac-$\alpha$ combinations. 
Similarly, Sagnac-$\beta$ and Sagnac-$\gamma$ combinations can be obtained, respectively, by rotating the indices of the constraints $2\rightarrow3,3\rightarrow1$ and $2\rightarrow1,3\rightarrow2$. 

\subsection{Beacon-type combinations}
From the specific forms of the modified first-generation Beacon-P combination
\begin{equation}
	\begin{aligned}
		{P}_{1}(t) &=D_{\bar{2}^{'}}\eta_{1}+D_{\bar{2}^{'}3} \eta_{2^{\prime}}+D_{\bar{2}^{'}33^{\prime}} \eta_{1^{\prime}}+D_{\bar{1}} \eta_{2} \\
		&-\left(D_{\bar{1}}\eta_{2^{\prime}}+D_{\bar{1}3^{\prime}} \eta_{1}+D_{\bar{1}3^{\prime}3} \eta_{2}+D_{\bar{2}^{'}} \eta_{1^{'}}\right),
	\end{aligned}
\end{equation}
and the modified second-generation standard Beacon-P combination 
\begin{equation}
	\begin{aligned}
		P_{2}(t) &=\left(1-D_{2^{'}\bar{1}3^{'}31\bar{2}^{'}}+D_{33^{'}2^{'}\bar{1}3^{'}}-D_{2^{'}\bar{1}3^{'}}\right)\eta_{1}\\
		&+\left(-1+D_{33^{'}}+D_{2^{'}\bar{1}3^{'}31\bar{2}^{'}}-D_{33^{'}2^{'}\bar{1}3^{'}31\bar{2}^{'}}\right)\eta_{1^{'}}\\
		&+\left(D_{33^{'}2^{'}\bar{1}3^{'}3}-D_{33^{'}2^{'}\bar{1}}-D_{2^{'}\bar{1}3^{'}3}+D_{2^{'}\bar{1}}\right)\eta_{2} \\
		&+\left(D_{3}+D_{33^{'}2^{'}\bar{1}}-D_{2^{'}\bar{1}}-D_{2^{'}\bar{1}3^{'}31\bar{2}^{'}3}\right)\eta_{2^{'}},
	\end{aligned}
\end{equation}
it is observed that the coefficients satisfy the following relations
\begin{equation}\label{beacon}
	q_{3}=0,q_{3^{'}}=0.
\end{equation}
By substituting Eq.~\eqref{beacon} into Eq.~\eqref{TDIeq}, one finds
\begin{equation}
	q_{1}(1-D_{33^{'}})+q_{1^{'}}(1-D_{2^{'}\bar{1} 3^{'}})=0.
\end{equation}
Again, the above equation is nothing but Eq.~\eqref{nMTDIgen} by identifying $a=D_{33^{'}},b=D_{2^{'}\bar{1} 3^{'}}$.
Since there is a time-advance operator $D_{\bar{1}}$ in $b$, one applies the process described in Sec.~\ref{sec4.2} to systematically enumerate the relevant commutators related to the Beacon-P type combinations.
One possible choice is $\Delta=[ba\bar{b},a]=[D_{2^{'}\bar{1}3^{'}31\bar{2}^{'}},D_{33^{'}}]$, whose the coefficients $q_1,q_{1^{'}}$,
\begin{equation}
	\begin{aligned}
		q_{1} &= (1-D_{2^{'}\bar{1}3^{'}}-D_{2^{'}\bar{1}3^{'}31\bar{2}^{'}}+D_{33^{'}2^{'}\bar{1}3^{'}}),\\
		q_{1^{'}} &= -(1-D_{3{3}^{'}}-D_{2^{'}\bar{1}3^{'}31\bar{2}^{'}}+D_{33^{'}2^{'}\bar{1}3^{'}31\bar{2}^{'}}),
	\end{aligned}
\end{equation}
are recognized as those of the modified second-generation standard Beacon-P combination given above.

On the other hand, for the case of $\Delta=[b,ab\bar{a}]$, we arrive at an alternative Beacon combination
\begin{equation}
	\begin{aligned}
		&{P}({t})=\left(1-D_{2^{'}\bar{1}3^{'}}-D_{33^{\prime}2^{'}\bar{1}\bar{3}}+D_{2^{'}\bar{1}3^{'}33^{'}2^{'}\bar{1}\bar{3}}\right)\eta_{1} \\
		&+\left(-1+D_{33^{'}}-D_{{2}^{\prime}\bar{1}3^{\prime}33^{'}}+D_{33^{'}2^{'}\bar{1}\bar{3}}\right)\eta_{1^{'}} \\
		&+\left({D}_{{2}^{\prime}\bar{1}}-D_{33^{\prime}{2}^{\prime}\bar{1}}+D_{{2}^{\prime}\bar{1}3^{'}33^{'}2^{'}\bar{1}}-D_{33^{\prime}2^{'}\bar{1}\bar{3}2^{'}\bar{1}}\right) \eta_{2} \\
		&+\left(D_{3}-D_{2^{'}\bar{1}}-D_{2^{'}\bar{1}3^{'}3}+D_{33^{'}2^{'}\bar{1}\bar{3}2^{'}\bar{1}}\right)\eta_{2^{\prime}}.
	\end{aligned}
\end{equation}
It is readily verified that the commutators $[ab^2,bab]$, $[a\bar{b},\bar{b}a]$, $[aab,aba]$, among others, give rise to a class of higher-order (from the perspective of the geometric TDI) Beacon-P combinations. 
Similarly, Beacon-Q and Beacon-R combinations can be obtained, respectively, by rotating the indices of the constraints $3\rightarrow2,3^{'}\rightarrow2^{'}$ and $3\rightarrow1,3^{'}\rightarrow1{'}$.

\subsection{Relay-type combinations}
From the specific forms of the modified first-generation Relay-U combination
\begin{equation}
	\begin{aligned}
		{U}_{1}(t) &=\eta_{1}+D_{3} \eta_{2^{\prime}}+D_{33^{\prime}} \eta_{1^{\prime}}+D_{33^{\prime} 2^{\prime}} \eta_{3^{\prime}} \\
		&-\left(\eta_{1^{\prime}}+D_{2^{\prime}} \eta_{3^{\prime}}+D_{2^{\prime} 1^{\prime}} \eta_{2^{\prime}}+D_{2^{\prime} 1^{\prime} 3^{\prime}} \eta_{1}\right),
	\end{aligned}
\end{equation}
and the modified second-generation standard Relay-U combination 
\begin{equation}
	\begin{aligned}
		{U}_{2}(t) &=\left(D_{33^{\prime} 2^{\prime} 1^{\prime}\overline{3}}-1\right) \left(\eta_{1^{\prime}}+D_{2^{\prime}} \eta_{3^{\prime}}+D_{2^{\prime} 1^{\prime}} \eta_{2^{\prime}} +D_{2^{'}1^{'}3^{'}} \eta_{1}\right) \\
		&-\left(D_{2^{\prime} 1^{\prime} 3^{\prime}}-1\right)D_{3}\left(\eta_{2^{'}}+D_{3^{'}} \eta_{1^{\prime}}+D_{3^{\prime}2^{'}} \eta_{3^{\prime}}\right)\\
		&-\left(D_{33^{\prime} 2^{\prime} 1^{\prime}\overline{3}}-1\right)\eta_{1},
	\end{aligned}
\end{equation}
it is observed that the coefficients satisfy the following relations
\begin{equation}\label{relay}
	q_{2}=0,q_{3}=0.
\end{equation}
By substituting Eq.~\eqref{relay} into Eq.~\eqref{TDIeq}, we have
\begin{equation}
	q_{2^{'}}(D_{\bar{3}}-D_{3^{'}})+q_{3^{'}}(D_{\bar{2}^{'}}-D_{1^{'}\bar{3}})=0.
\end{equation}
The above equation is essentially Eq.~\eqref{nMTDIgen} by recognizing $\alpha=q_{2'}D_{\bar{3}}, \beta=q_{3^{'}}D_{\bar{2}^{'}}$, $a=D_{33^{'}}$, and $b=D_{2^{'}1^{'}\bar{3}}$.
Since there is a time-advance operator $D_{\bar{3}}$ in $b$, one applies the algorithm described in Sec.~\ref{sec4.2} to systematically enumerate the relevant commutators for the Relay-U type combinations.
The simplest choice is $\Delta=[ba,ab]=[D_{\bar{1}3^{'}31},D_{\bar{2}^{'}33^{'}2^{'}}]$, whose the coefficients $q_{2^{'}},q_{3^{'}}$,
\begin{equation}
	\begin{aligned}
		&{q}_{2^{\prime}}=\left(1-D_{2^{\prime} 1^{\prime} \overline{3}}-D_{2^{\prime} 1^{\prime} 3^{\prime}}+D_{33^{\prime} 2^{\prime} 1^{\prime} \overline{3} 2^{\prime} 1^{\prime} \overline{3}}\right) D_{3}, \\
		&{q}_{3^{\prime}}=-\left(1-D_{33^{\prime}}-D_{33^{\prime} 2^{\prime} 1^{\prime} \overline{3}}+D_{2^{\prime} 1^{\prime} 3^{\prime} 33^{\prime}}\right) D_{2^{\prime}},
	\end{aligned}
\end{equation}
are recognized as those of the modified second-generation Relay-U combination given above.

On the other hand, for the case of $\Delta=[ba\bar{b},a]$, we arrive at an alternative Relay combination
\begin{equation}
	\begin{aligned}
		{U}({t})&=\left(1-{D}_{2^{\prime} 1^{\prime} 3^{\prime}}-{D}_{2^{\prime} 1^{\prime} 3^{\prime} 3 \bar{1}^{'} \bar{2}^{'}}+D_{33^{\prime} 2^{\prime} 1^{\prime} 3^{\prime}}\right) \eta_{1} \\
		&-\left(1-D_{33^{\prime}}-D_{2^{\prime} 1^{\prime} 3^{\prime} 3 \bar{1}^{'} \bar{2}^{'}}+D_{33^{\prime} 2^{\prime} 1^{\prime} 3^{\prime} 3 \bar{1}^{'} \bar{2}^{'}}\right) \eta_{1^{\prime}} \\
		&+\left(1-D_{2^{\prime} 1^{\prime} \overline{3}}+D_{33^{\prime} 2^{\prime} 1^{\prime} \overline{3}}-D_{2^{\prime} 1^{\prime} 3^{\prime} 3 \bar{1}^{'} \bar{2}^{'}}\right) D_{3} \eta_{2^{\prime}} \\
		&-\left(1-D_{33^{\prime}}-D_{2^{\prime} 1^{\prime} 3^{\prime} 3 \bar{1}^{'} \bar{2}^{'}}+D_{33^{\prime} 2^{\prime} 1^{\prime} 3^{\prime} 3 \bar{1}^{'} \bar{2}^{'}}\right) D_{2^{\prime}} \eta_{3^{\prime}}.
	\end{aligned}
\end{equation}
It is readily verified that the commutators $[ab^2,bab]$, $[a\bar{b},\bar{b}a]$, $[aab,aba]$, among others, give rise to a class of higher-order (from the perspective of the geometric TDI) Relay-U combinations. 
Similarly, Relay-V, Relay-W combinations can be obtained, respectively, by rotating the indices of the constraints $2\rightarrow3,3\rightarrow1$ and $2\rightarrow1,3\rightarrow2$.

\section{The specific forms of the lowest-order commutators}\label{C}

In this Appendix, we enumerate the forms of the lowest-order commutators for our proposed algorithm.
Using the notations in Sec.~\ref{sec4.2}, we denote the length of the commutator by $2n$.
Due to the presence of the time-advance operators, it is noted that the two terms of the commutator do not necessarily have the same length.
For $n=2$, the total count of $a$'s, $\bar{a}$'s, $b$'s, and $\bar{b}$'s is $2n=4$.
This implies two possibilities, $4=1+3$ or $4=2+2$, such as $[ab,ba]$ and $[a,ba\bar{b}]$. 
By taking $n=2$ and $n=3$ as examples, we enumerate all possible forms of the relevant commutators that satisfy Eq.~\eqref{permuXY} for the proposed algorithm in Tab.~\ref{t1}.
It is noted that even at the lowest order, the number of possible commutators is much more significant compared to the original algorithm presented in Sec.~\ref{sec3}. 

	\begin{table*}[]
		\centering
		\caption{The specific forms of the commutators with $n=2$ and $n=3$}
		\label{t1}
		\newcommand{\tabincell}[2]{\begin{tabular}{@{}#1@{}}#2\end{tabular}}
		\renewcommand\arraystretch{2}
		\begin{tabular}{ccc}
			\hline
			\hline
			Length & & All possible relevant commutators \\ \hline
			\multirow{3}{*}{$2n=4$}      & \multirow{2}{*}{1+3}     & $[a,ba\bar{b}]$ $[a,\bar{b}ab]$ $[ab\bar{a},b]$ $[a\bar{b}\bar{a},\bar{b}]$                 \\
			& 							& $[b, \bar{a} b a]$ $[b \bar{a} \bar{b}, \bar{a}]$ $[\bar{a}, \bar{b} \bar{a} b]$ $[\bar{a} \bar{b} a, \bar{b}]$	\\
			\cline{2-3}
			& \multirow{1}{*}{2+2}    & $[ab,ba]$ $[a\bar{b},\bar{b}a]$ $[b\bar{a},\bar{a}b]$ $[\bar{a}\bar{b},\bar{b}\bar{a}]$                 \\
			\cline{1-3}
			\multirow{19}{*}{$2n=6$} & \multirow{8}{*}{1+5} & $[a, a a b \bar{a} \bar{b}]$ $[a, a a \bar{b} \bar{a} b]$ $[a, a b a \bar{b} \bar{a}]$ $[a, a b \bar{a} \bar{b} a]$ $[a, a \bar{b} a b \bar{a}]$ $[a, a \bar{b} \bar{a} b a]$ $[a,baa\bar{b}\bar{a}]$ $[a,bba\bar{b}\bar{b}]$                \\
			&                      & $[a,b\bar{a}\bar{b}aa]$ $[a,\bar{a}baa\bar{b}]$ $[a,\bar{a}ba\bar{b}a]$ $[a,\bar{a}\bar{b}aab]$ $[a,\bar{a}\bar{b}aba]$ $[a,\bar{b}aab\bar{a}]$ $[a,\bar{b}\bar{a}baa]$ $[a,\bar{b}\bar{b}abb]$                \\
			&                      &$[aab\bar{a}\bar{a},b]$ $[a a \bar{b} \bar{a} \bar{a}, \bar{b}]$ $[a b b \bar{a} \bar{b}, b]$ $[a b \bar{a} \bar{a} \bar{b}, \bar{a}]$ $[a b \bar{a} \bar{b} \bar{a}, \bar{a}]$ $[a b \bar{a} \bar{b} \bar{b}, \bar{b}]$ $[a \bar{b} \bar{a} b b, b]$ $[a \bar{b} \bar{a} b \bar{a}, \bar{a}]$                 \\
			&                      & $[a \bar{b} \bar{a} \bar{a} b, \bar{a}]$ $[a \bar{b} \bar{b} \bar{a} b, \bar{b}]$ $[b, b a b \bar{a} \bar{b}]$ $[b, b a \bar{b} \bar{a} b]        $ $[b, b b a \bar{b} \bar{a}]$ $[b, b b \bar{a} \bar{b} a]$ $[b, b \bar{a} b a \bar{b}]$ $[b, b \bar{a} \bar{b} a b]$          \\
			&                      & $[b, \bar{a} b b a \bar{b}]$ $[b, \bar{a} \bar{a} b a a]$ $[b, \bar{a} \bar{b} a b b]$ $[b, \bar{b} a b b \bar{a}]$ $[b, \bar{b} a b \bar{a} b]$ $[b, \bar{b} \bar{a} b a b]$ $[b, \bar{b} \bar{a} b b a]$ $[ba\bar{b}\bar{a}\bar{a},\bar{a}]$                  \\
			&                      & $[b a \bar{b} \bar{a} \bar{b}, \bar{b}]$ $[b a \bar{b} \bar{b} \bar{a}, \bar{b}]$ $[b b \bar{a} \bar{b} \bar{b}, \bar{a}]$ $[b \bar{a} \bar{a} \bar{b} a, \bar{a}]$ $[b \bar{a} \bar{b} a \bar{b}, \bar{b}]$ $[b \bar{a} \bar{b} \bar{b} a, \bar{b}]$ $[\bar{a}, \bar{a} b a \bar{b} \bar{a}]$ $[\bar{a}, \bar{a} b \bar{a} \bar{b} a]$                 \\
			&                      & $[\bar{a},\bar{a}\bar{a}ba\bar{b}]$ $[\bar{a}, \bar{a} \bar{a} \bar{b} a b]$ $[\bar{a}, \bar{a} \bar{b} a b \bar{a}]$ $[\bar{a}, \bar{a} \bar{b} \bar{a} b a]$ $[\bar{a}, \bar{b} a b \bar{a} \bar{a}]$ $[\bar{a}, \bar{b} \bar{a} \bar{a} b a]$ $[\bar{a}, \bar{b} \bar{b} \bar{a} b b]$ $[\bar{a} b a \bar{b} \bar{b}, \bar{b}]$                 \\
			&                      & $[\bar{a} \bar{a} \bar{b} a a, \bar{b}]$ $[\bar{a} \bar{b} \bar{b} a b, \bar{b}]$ $[\bar{b}, \bar{b} a b \bar{a} \bar{b}]$ $[\bar{b}, \bar{b} a \bar{b} \bar{a} b]$ $[\bar{b}, \bar{b} \bar{a} b a \bar{b}]$ $[\bar{b}, \bar{b} \bar{a} \bar{b} a b]$ $[\bar{b}, \bar{b} \bar{b} a b \bar{a}]$ $[\bar{b}, \bar{b} \bar{b} \bar{a} b a]$                 \\
			\cline{2-3}
			& \multirow{7}{*}{2+4} & $[a a, a b a \bar{b}]$ $[a a, a \bar{b} a b]$ $[a a, b a a \bar{b}]$ $[a a, b a \bar{b} a]$ $[a a, \bar{b} a a b]$ $[a a, \bar{b} a b a]$ $[a a b \bar{a}, a b]$ $[a a b \bar{a}, b a]$                  \\
			&                      & $[a a \bar{b} \bar{a}, a \bar{b}]$ $[a a \bar{b} \bar{a}, \bar{b} a]$ $[a b, b b a \bar{b}]$ $[a b, \bar{a} b a a]$ $[a b, \bar{b} a b b]$ $[a b b \bar{a}, b b]$ $[a b \bar{a} b, b b]$ $[a b \bar{a} \bar{a}, b \bar{a}]$                  \\
			&                       & $[a b \bar{a} \bar{a}, \bar{a} b]$ $[a \bar{b}, b a \bar{b} \bar{b}]$ $[a \bar{b}, \bar{a} \bar{b} a a]$ $[a \bar{b}, \bar{b} \bar{b} a b]$ $[a \bar{b} \bar{a} \bar{a}, \bar{a} \bar{b}]$ $[a \bar{b} \bar{a} \bar{a}, \bar{b} \bar{a}]$ $[a \bar{b} \bar{a} \bar{b}, \bar{b} \bar{b}]$ $[a \bar{b} \bar{b} \bar{a}, \bar{b} \bar{b}]$                \\
			&                      & $[b a, b b a \bar{b}]$ $[b a, \bar{a} b a a]$ $[b a, \bar{b} a b b]$ $[b a b \bar{a}, b b]$ $[b a \bar{b} \bar{b}, \bar{b} a]$ $[b b, b \bar{a} b a]$ $[b b, \bar{a} b a b]$ $[b b, \bar{a} b b a]$                 \\
			&                      & $[b b \bar{a} \bar{b}, b \bar{a}]$ $[b b \bar{a} \bar{b}, \bar{a} b]$ $[b \bar{a}, \bar{a} \bar{a} b a]$ $[b \bar{a}, \bar{b} \bar{a} b b]$ $[b \bar{a} \bar{a} \bar{b}, \bar{a} \bar{a}]$ $[b \bar{a} \bar{b} \bar{a}, \bar{a} \bar{a}]$ $[b \bar{a} \bar{b} \bar{b}, \bar{a} \bar{b}]$ $[b \bar{a} \bar{b} \bar{b}, \bar{b} \bar{a}]$                 \\
			&                      & $[\bar{a}b,\bar{a}\bar{a}ba]$ $[\bar{a} b, \bar{b} \bar{a} b b]$ $[\bar{a} b \bar{a} \bar{b}, \bar{a} \bar{a}]$ $[\bar{a} \bar{a}, \bar{a} \bar{b} \bar{a} b]$ $[\bar{a} \bar{a}, \bar{b} \bar{a} b \bar{a}]$ $[\bar{a} \bar{a}, \bar{b} \bar{a} \bar{a} b]$ $[\bar{a} \bar{a} \bar{b} a, \bar{a} \bar{b}]$ $[\bar{a} \bar{a} \bar{b} a, \bar{b} \bar{a}]$                  \\
			&                      & $[\bar{a} \bar{b}, \bar{b} \bar{b} \bar{a} b]$ $[\bar{a} \bar{b} a a, \bar{b} a]$ $[\bar{a} \bar{b} a \bar{b}, \bar{b} \bar{b}]$ $[\bar{a} \bar{b} \bar{b} a, \bar{b} \bar{b}]$ $[\bar{b} a, \bar{b} \bar{b} a b]$ $[\bar{b} a \bar{b} \bar{a}, \bar{b} \bar{b}]$ $[\bar{b} \bar{a}, \bar{b} \bar{b} \bar{a} b]$ $[\bar{b} \bar{a} \bar{b} a, \bar{b} \bar{b}]$                \\
			\cline{2-3}
			& \multirow{4}{*}{3+3} & $[a a b, a b a]$ $[a a b, b a a]$ $[a a \bar{b}, a \bar{b} a]$ $[a a \bar{b}, \bar{b} a a]$ $[a b a, b a a]$ $[a b b, b a b]$ $[a b b, b b a]$ $[a b \bar{a}, \bar{a} b a]$                  \\
			&                      & $[a \bar{b} a, \bar{b} a a]$ $[a \bar{b} \bar{a}, \bar{a} \bar{b} a]$ $[a \bar{b} \bar{b}, \bar{b} a \bar{b}]$ $[a \bar{b} \bar{b}, \bar{b} \bar{b} a]$ $[b a b, b b a]$ $[b a \bar{b}, \bar{b} a b]$ $[b b \bar{a}, b \bar{a} b]$ $[b b \bar{a}, \bar{a} b b]$                 \\
			&                      & $[b \bar{a} b, \bar{a} b b]$ $[b \bar{a} \bar{a}, \bar{a} b \bar{a}]$ $[b \bar{a} \bar{a}, \bar{a} \bar{a} b]$ $[b \bar{a} \bar{b}, \bar{b} \bar{a} b]$ $[\bar{a} b \bar{a}, \bar{a} \bar{a} b]$ $[\bar{a} \bar{a} \bar{b}, \bar{a} \bar{b} \bar{a}]$ $[\bar{a} \bar{a} \bar{b}, \bar{b} \bar{a} \bar{a}]$ $[\bar{a} \bar{b} \bar{a}, \bar{b} \bar{a} \bar{a}]$                  \\
			&                      & $[\bar{a} \bar{b} \bar{b}, \bar{b} \bar{a} \bar{b}]$ $[\bar{a} \bar{b} \bar{b}, \bar{b} \bar{b} \bar{a}]$ $[\bar{b} a \bar{b}, \bar{b} \bar{b} a]$ $[\bar{b} \bar{a} \bar{b}, \bar{b} \bar{b} \bar{a}]$                 \\ \hline
		\end{tabular}
	\end{table*}

\section{Specific higher-order TDI combinations obtained using the lower-order commutators}\label{D}

This section elaborates on constructing specific higher-order TDI solutions based on the lower-order commutators.
We will be focused on two possible schemes.
The first feasible scheme is multiplying an arbitrary monomial to the left of a relevant commutator.
For instance, $a[ba,ab]$ is a valid commutator satisfying Eq.~\eqref{permuXY}, which is obtained by multiplying $a$ to the left of $[ba,ab]$. 
Similarly, one may derive a higher-order commutator by multiplying an additional monomial to the right of a valid commutator.
However, as will be shown shortly, the second choice implies some subtlety in deriving the corresponding TDI coefficients.

For the first scheme, it is not difficult to derive the corresponding TDI coefficients defined in Eq.~\eqref{nMTDI}. 
Taking $a[ba, ab]$ as an example, one only needs to multiply to the left of the existing TDI solution, Eq.~\eqref{M16}, by the monomial in question, $a$, namely
\begin{equation}\label{aBBB}
	\begin{aligned}
		{\tilde q}_{1}&=a{q}_{1}=a(1-b-b a+a b^{2}), \\
		{\tilde q}_{1^{\prime}}&=a{q}_{1^{\prime}}=-a\left(1-a-a b+b a^{2}\right) .
	\end{aligned}
\end{equation}
The validity and generality of the above scheme can be verified straightforwardly by either substituting Eq.~\eqref{aBBB} back into Eq.~\eqref{nMTDI} or closely observing the process discussed in Sec.~\ref{sec3.2}.

For the second scheme, although the validity of the commutator is apparent, the corresponding TDI coefficients do not possess a simple relation with the lower-order counterpart. 
We first consider a simple example, $[ba,ab]b$. 
It is not difficult to show that the TDI coefficients of Eq.~\eqref{nMTDI} for this example is 
\begin{equation}
	\begin{aligned}
			{\tilde q}_{1}&=1-b-b a+a b^{2},\\
		{\tilde q}_{1^{\prime}}&=-\left(1-a-a b+b a^{2}+[b a, a b]\right) ,
	\end{aligned}
\end{equation} 
which does not have a simple relation with the TDI solution of the original commutator $[ba, ab]$, Eq.~\eqref{M16}.

However, the above result motivates us to assume an {\it ansatz} for the TDI solution and then solve for its specific form.
To this end, we will illustrate the approach by a more complicated example, $[ab^2, bab]b\bar{a}$.
We utilize the following ansatz
\begin{equation}\label{BBBa}
	\begin{aligned}
		{\tilde q}_{1} &={q}_{1}+\left[a b^{2}, b a b\right] \alpha, \\
		{\tilde q}_{1^{\prime}} &=q_{1^{\prime}}+\left[a b^{2}, b a b\right] \beta,
	\end{aligned}
\end{equation}
where $\alpha$ and $\beta$ are operator polynomials to be determined, $q_1$ and $q_{1^{'}}$ are the TDI solution of Eq.~\eqref{nMTDI} associated with the commutator $[ab^2,bab]$. 
Substitute Eq.~\eqref{BBBa} back into Eq.~\eqref{nMTDI}, one finds an equation
\begin{equation}\label{r}
	\left[a b^{2}, b a b\right](1+\alpha(1-a)+\beta(1-b)) ,
\end{equation}
which should be nothing but $[ab^2,bab]\bar{b}a$.
This implies 
\begin{equation}\label{alphabetaeq}
	\alpha(1-a)+\beta(1-b)=\bar{b} a -1.
\end{equation}

Solving Eq.~\eqref{alphabetaeq} for $\alpha$ and $\beta$ will provide us with the desired TDI coefficients.
Employing a similar trick in the derivation of Eq.~\eqref{newMTDI}, we have
\begin{equation}
\bar{b} a=-\bar{b}(1-a)+\bar{b}=-\bar{b}(1-a)+\bar{b}(1-b)+1 ,
\end{equation}
in other words,
\begin{equation}
\bar{b} a-1=(-\bar{b})(1-a)+\bar{b}(1-b) .
\end{equation}
The simplest choice (that is, the lowest-order solution) is therefore $-\alpha=\beta=\bar{b}$, and subsequently, we have
\begin{equation}\label{[]b}
	\begin{aligned}
		{\tilde q}_{1} &=q_{1}-\left[a b^{2}, b a b\right]\bar{b}, \\
		{\tilde q}_{1^{\prime}} &=q_{1^{\prime}}+\left[a b^{2}, b a b\right]\bar{b} .
	\end{aligned}
\end{equation}
We observe that the above procedure is, by and large, general.
It provides an approach to construct two specific classes of higher-order (from the geometric TDI perspective) TDI solutions from the lower-order ones.

\section{Sensitivity funtions for Sagnac and novel Sagnac-inspired combinations}\label{E}

This Appendix presents the expressions for the averaged response functions of gravitational waves and noise power spectral density for Sagnac-type combinations elaborated in the main text.
Here, we consider that the residual noise primarily comprises the shot noise and test mass vibrations, whereas the other ones, such as the clock jitter noise, are suppressed below the noise floor. 

For the Sagnac combination, the noise power spectral density $N_{\alpha}(u)$ can be expressed as
	\begin{equation}\label{c1}
		N_{\alpha}(u)=\left[16\frac{s_{a}^{2}L^{2}}{c^{4}u^{2}}\Big(3-2\mathrm{cos}u-2\mathrm{cos}3u\Big)+24\frac{s_{x}^{2}u^{2}}{L^{2}}\right]\mathrm{sin}^{2}\frac{3u}{2},
	\end{equation}
and the gravitational waves averaged response function $R_{\alpha}(u)$ reads
	\begin{equation}\label{c2}
		\begin{aligned}
			R_{\alpha}(u)=\frac{1}{6}\mathrm{sin}^2\frac{3u}{2}&\left[28-24\mathrm{cos}u-4\mathrm{cos}3u+48\mathrm{log}2\left(1+2\mathrm{cos}u\right)+72\left(2\mathrm{log}2-\mathrm{log}3\right)\left(2+\mathrm{cos}3u\right)\right.\\
			&\left.+\frac{-204\mathrm{sin}u+147\mathrm{sin}2u-30\mathrm{sin}3u}{u}+\frac{-81+144\mathrm{cos}u-81\mathrm{cos}2u+18\mathrm{cos}3u}{u^2}\right.\\
			&\left.+\frac{-108\mathrm{sin}u+81\mathrm{sin}2u-18\mathrm{sin}3u}{u^3}+48\left(4\mathrm{Ci}u-7\mathrm{Ci}2u+3\mathrm{Ci}3u\right)+96\mathrm{cos}u\left(\mathrm{Ci}u-\mathrm{Ci}2u\right)\right.\\
			&\left.+72\mathrm{cos}3u\left(\mathrm{Ci}u-2\mathrm{Ci}2u+\mathrm{Ci}3u\right)+72\mathrm{sin}3u\left(\mathrm{Si}u-2\mathrm{Si}2u+\mathrm{Si}3u\right)\right],
		\end{aligned}
	\end{equation}
where $u=2\pi fc/L$ and $f$ is observed frequency, SinIntergral $\mathrm{Si}\left(z\right)=\int_{0}^{z}\mathrm{sin}t/tdt$ and CosIntegral $\mathrm{Ci}\left(z\right)=-\int_{z}^{\infty}\mathrm{cos}t/tdt$.

For the Sagnac-inspired combination, the noise power spectral density $N_{s}(u)$ can be expressed as
\begin{equation}\label{c3}
	\begin{aligned}
		N_{s}(u)=&64\left(4+3\mathrm{cos}u+2\mathrm{cos}2u+3\mathrm{cos}3u+2\mathrm{cos}4u+\mathrm{cos}5u+2\mathrm{cos}6u+\mathrm{cos}7u\right)\mathrm{sin}^4\frac{u}{2}\times\\
		&\left(\frac{2s_{a}^{2}L^{2}}{c^4u^{2}}\left(13+\mathrm{cos}u-8\mathrm{cos}2u-4\mathrm{cos}3u-2\mathrm{cos}4u\right)+\frac{s_{x}^{2}u^{2}}{L^{2}}\left(16+10\mathrm{cos}u-2\mathrm{cos}2u+2\mathrm{cos}3u+\mathrm{cos}4u\right)\right),
	\end{aligned}
\end{equation}
and the averaged response function of gravitational waves $R_{s}(u)$ is found to be
\begin{equation}\label{c4}
	\begin{aligned}
		R_{s}(u)=&\frac{2}{3}\left(4+3\mathrm{cos}u+2\mathrm{cos}2u+3\mathrm{cos}3u+2\mathrm{cos}4u+\mathrm{cos}5u+2\mathrm{cos}6u+\mathrm{cos}7u\right)\mathrm{sin}^4\frac{u}{2}\times\\
		&\left[96\left(41+31\mathrm{cos}u+4\mathrm{cos}2u+22\mathrm{cos}3u+10\mathrm{cos}4u\right)\mathrm{log}2-144\left(11+6\mathrm{cos}u+7\mathrm{cos}3u+3\mathrm{cos}4u\right)\mathrm{log}3\right.\\
		&\left.+944\mathrm{sin}^2\frac{u}{2}+64\mathrm{sin}^{2}\frac{u}{2}\left(15\mathrm{cos}u+3\mathrm{cos}2u+\mathrm{cos}3u\right)-\frac{3\left(684\mathrm{sin}u-273\mathrm{sin}2u-146\mathrm{sin}3u+86\mathrm{sin}4u-10\mathrm{sin}5u+\mathrm{sin}6u\right)}{u}\right.\\
		&\left.+\frac{12\mathrm{sin}^2\frac{u}{2}\left(-16+78\mathrm{cos}u-29\mathrm{cos}3u+16\mathrm{cos}4u+5\mathrm{cos}5u\right)}{u^{2}}-\frac{24\mathrm{sin}^3\frac{u}{2}\left(96\mathrm{cos}\frac{u}{2}-50\mathrm{cos}\frac{3u}{2}-18\mathrm{cos}\frac{5u}{2}+21\mathrm{cos}\frac{7u}{2}+5\mathrm{\frac{9u}{2}}\right)}{u^{3}}\right.\\
		&\left.+48\left(49\mathrm{Ci}u-82\mathrm{Ci}2u+33\mathrm{Ci}3u\right)+96\mathrm{cos}u\left(22\mathrm{Ci}u-31\mathrm{Ci}2u+9\mathrm{Ci}3u\right)+384\mathrm{cos}2u\left(\mathrm{Ci}u-\mathrm{Ci}2u\right)\right.\\
		&\left.+48\mathrm{cos}3u\left(23\mathrm{Ci}u-44\mathrm{cos}2u+21\mathrm{Ci}3u\right)+48\mathrm{cos}4u\left(11\mathrm{Ci}u-20\mathrm{Ci}2u+9\mathrm{Ci}3u\right)\right.\\
		&\left.+432\left(-\mathrm{sin}u+2\mathrm{sin}3u+\mathrm{sin}4u\right)\left(\mathrm{Si}u-2\mathrm{Si}2u+\mathrm{Si}3u\right)\right].
		\end{aligned}
\end{equation}

The resultant sensitivity function is
\begin{equation}\label{c5}
	S(u)=\frac{\sqrt{N(u)}}{\sqrt{\frac{2}{5}}\sqrt{R(u)}}.
\end{equation}
Fig.~\ref{fig3} and Fig.~\ref{fig4} are evaluated using Eqs.~\eqref{c1}$-$\eqref{c5}.

\bibliographystyle{h-physrev}
\bibliography{ref}

\end{document}